\documentclass[a4paper,11pt]{article}
\usepackage{jcappub}
\usepackage{amsmath}
\usepackage{graphicx}
\usepackage{latexsym}
\usepackage{xspace}
\usepackage{xcolor}
\usepackage{bm}
\usepackage{relsize}
\usepackage{tabularx}
\usepackage{multirow}
\usepackage{amssymb}
\usepackage[table]{xcolor}
\usepackage{braket}
\usepackage{comment}
\usepackage[utf8]{inputenc}
\usepackage{slashed}
\usepackage{empheq}
\usepackage{subfigure}
\usepackage{here}

\usepackage[ruled,vlined]{algorithm2e}

\makeatletter
\gdef\@fpheader{}
\g@addto@macro\bfseries{\boldmath}
\makeatother

\newcommand{\prn}[1]{\left( {#1} \right)}
\newcommand{\com}[1]{\left[ {#1} \right]}

\newcommand{\dif}[2]{\frac{\mathrm{d} #1}{\mathrm{d} #2}}
\newcommand{\pdif}[2]{\frac{\partial #1}{\partial #2}}

\newcommand{\uIR}{\mathrm{IR}}
\newcommand{\uUV}{\mathrm{UV}}

\newcommand{\ie}{\textsl{i.e.~}}

\newcommand{\dd}{\mathrm{d}}

\newcommand{\sss}[1]{{\scriptscriptstyle{#1}}}

\newcommand{\uPl}{\mathrm{Pl}}

\newcommand{\usssPl}{\sss{\uPl}}

\newcommand{\mpl}{M_\usssPl}
\newcommand{\Mp}{M_\usssPl}

\newcommand{\efolds}{$e$-folds}

\newcommand{\beq}{\begin{equation}}
\newcommand{\eeq}{\end{equation}}
\newcommand{\bea}{\begin{equation}\begin{aligned}}
\newcommand{\eea}{\end{aligned}\end{equation}}

\newlength{\wsingfig}
\newlength{\wdblefig}
\newlength{\wquadfig}
\newlength{\wtriplefig}
\newcommand{\Fig}[1]{Fig.~{\ref{#1}}}
\newcommand{\Figs}[1]{Figs.~{\ref{#1}}}

\newcommand{\Sec}[1]{Sec.~\ref{#1}}

\newcommand{\App}[1]{Appendix~\ref{#1}}

\subheader{}

\title{Numerical simulation of the stochastic formalism including non-Markovianity}

\author[a,b]{Masahiro Kawasaki}
\author[c,d]{Tomotaka Kuroda,}

\affiliation[a]{ICRR, University of Tokyo, Kashiwa 277-8582, Japan}
\affiliation[b]{Kavli IPMU (WPI), UTIAS, University of Tokyo, Kashiwa 277-8583, Japan}
\affiliation[c]{Department of Physics, Institute of Science Tokyo Tokyo, 152-8551, Japan}
\affiliation[d]{Cosmology, Gravity and Astroparticle Physics Group, Center for Theoretical Physics of the Universe, Institute for Basic Sclence (IBS), Daejeon, 34126, Korea}

\emailAdd{kawasaki@icrr.u-tokyo.ac.jp}
\emailAdd{kuroda.t.ad@m.titech.ac.jp}

\date{today}

\begin{document}
\sloppy

\abstract{We numerically investigate stochastic dynamics in cosmology by solving Langevin equations for Infrared (IR) modes with stochastic noises generated by Ultraviolet (UV) modes at the coarse-graining scale. By construction, the stochastic formalism relies on the separation of scales, which requires solving the equations for UV modes on top of the evolving IR modes for all modes at every time step, leading to a non-Markovian system in general. In this paper, working on a de Sitter background, we analyze several representative models by simultaneously solving the Langevin equations for IR modes and the equations for UV modes at each time step. We demonstrate that once the effects of effective masses are treated consistently by our simulation, the flat direction in the minimal supersymmtric model (MSSM) does not saturate but instead evolves as an exactly flat direction. Furthermore, we investigate memory effects in simple two models; $V=\lambda\phi^4$ and $V=\mu\phi\chi + \lambda\phi^4$, and non-Markovian contributions can lead to quantitative differences, even in stationary configurations, when compared with Markovian approximations, particularly in the strong-coupling regime.}

\arxivnumber{}

\maketitle


\flushbottom


\section{Introduction}\label{Sec:1}

Cosmological inflation refers to a period of accelerated expansion in the very early universe. Perturbations generated as quantum fluctuations during inflation, in a spacetime close to de Sitter, provide a compelling explanation for observed phenomena such as the temperature anisotropies of the cosmic microwave background (CMB). Therefore, a thorough understanding of the quantum aspects of quasi de Sitter spacetimes, including radiative corrections, is essential for making reliable predictions of observable phenomena from theories derived from first principles. 

In order to describe such quantum perturbations, this paper is devoted to the stochastic formalism~\cite{Starobinsky:1986fx,Starobinsky:1994bd} as a theoretical framework. The stochastic formalism is formulated non-perturbatively~\cite{PerreaultLevasseur:2013kfq,Burgess:2014eoa,Burgess:2015ajz,Moss:2016uix,Grain:2017dqa,Gorbenko:2019rza,Baumgart:2019clc,Mirbabayi:2019qtx,Mirbabayi:2020vyt, Cohen:2020php,Cohen:2021fzf,Cruces:2018cvq,Cruces:2021iwq,Cruces:2022imf,Li:2025azq} and can be understood as an effective field theory for IR modes, incorporating quantum corrections from UV modes fluctuations. One important consequence of this framework is its ability to address the problem of IR divergences for light spectator fields in de Sitter spacetime, which is notorious in perturbative calculations. The stochastic formalism can resum the corresponding IR loop contributions and interprets the resulting secular growth as an increase in statistical variances~\cite{Tsamis:2005hd,Enqvist:2008kt,Seery:2010kh,Garbrecht:2013coa,Garbrecht:2014dca,Tokuda:2017fdh,Tokuda:2018eqs,Honda:2023unh,Cespedes:2023aal,Palma:2023idj,Palma:2023uwo, Palma:2025oux}. Another technical advantage of the stochastic formalism is that probability distribution functions (PDFs), including information from correlation functions of arbitrarily high order, can be computed directly as outputs of the formalism~\cite{Markkanen:2019kpv,Markkanen:2020bfc,Cable:2020dke,Cable:2022uwd,Cable:2023gdz,Bhattacharya:2022wjl}, whereas perturbation theories in practice require truncation at some finite order in general. 

In addition, the stochastic formalism has been widely employed to describe cosmological perturbations in non-linear and even non-perturbative regimes. For instance, phenomena such as primordial black holes (PBHs) or scalar induced gravitational waves are triggered by large-amplitude fluctuations that necessitate descriptions beyond conventional perturbation theory. In this direction, the stochastic-$\delta N$ formalism~\cite{Fujita:2013cna, Fujita:2014tja, Vennin:2015hra} has been developed, enabling a non-perturbative description of curvature perturbations as differences in the local amounts of inflation, quantified by the number of inflationary \efolds\ via the $\delta N$ formalism as a geometrical relationship~\cite{Starobinsky:1982ee, Starobinsky:1986fxa, Sasaki1996, Sasaki:1998ug, Lyth:2004gb}. From a technical perspective, the statistics of that number of \efolds~can be computed by solving a first-passage-time problem~\cite{Vennin:2015hra,Pattison:2017mbe,Ezquiaga:2019ftu}, which can give rise to strongly non-Gaussian, heavy-tailed distributions. Consequently, the stochastic-$\delta N$ formalism has been applied to various models related to PBHs~\cite{Ballesteros:2020sre, Figueroa:2020jkf, Figueroa:2021zah,Tomberg:2023kli,Mishra:2023lhe,Tada:2023fvd, Mizuguchi:2024kbl,Murata:2025onc,Kuroda:2025coa,Raatikainen:2025gpd}, where these heavy tails significantly enhance probabilities to exceed the PBH formation threshold. Such probabilities can be evaluated using various techniques~\cite{Ando:2020fjm, Tada:2021zzj, Kitajima:2021fpq, Hooshangi:2021ubn, Gow:2022jfb, Raatikainen:2023bzk, Animali:2024jiz, Vennin:2024yzl, Jackson:2024aoo, Animali:2025pyf, Choudhury:2025kxg}. We note, however, that most previous studies, such as the first passage time analysis, rely on the assumption that the system is Markovian, even though non-Markovianity could generically arise, as mentioned below. 

One formulation of the stochastic formalism is based on the Schwinger-Keldysh path integral approach for open systems~\cite{Morikawa:1989xz,PerreaultLevasseur:2013kfq,Moss:2016uix,Pinol:2020cdp, Panda:2025tpu}, in which UV modes are treated as an environment that influences the system of IR modes through stochastic diffusion by UV modes at the coarse-graining scale. In this derivation, the dynamics of IR modes are described by Langevin equations with statistical noise terms. These noise terms are determined by the UV modes evolving on top of the IR background, which in principle requires solving not only the Langevin equations for IR modes but also the equations for all scales of UV modes at each time step~\cite{Figueroa:2020jkf,Figueroa:2021zah,Launay:2025kef,Launay:2025lnc}. Moreover, such coupled systems of stochastic differential equations could exhibit non-Markovian behavior~\cite{Cruces:2022imf,Figueroa:2021zah,Pinol:2020cdp,Tomberg:2022mkt,Tomberg:2024evi,Brahma:2024ycc,Calderon-Figueroa:2025dto}, as the evolution generally depends on the entire history of the system. (See \Sec{Sec:2} for details).
 
In this work, we perform numerical simulations of the stochastic formalism in its fully general implementation, in which the dynamical equations of both IR and UV modes are solved simultaneously, as discussed above. The numerical codes, demonstrated in detail in \App{Sec:A}, are applied to several representative models. As a first application, we consider a potential constructed by a flat direction coupled to non-flat directions in the context of the minimal supersymmetric standard model (MSSM)~\cite{Enqvist:2003gh,Allahverdi:2006we,Gherghetta:1995dv}. We find that the flat direction is not saturated, but instead continues to grow along the flat direction. This behavior arises because our numerical framework consistently captures the contributions of the effective masses generated through interactions with the non-flat directions. As further applications, we investigate non-Markovian effects within the stochastic dynamics. Our numerical results demonstrate the presence of memory effects for all fields considered, which could modify the equilibrium probability distributions. 

The remainder of this paper is organized as follows. In \Sec{Sec:2}, we review the derivation of the stochastic formalism and discuss the properties of the resulting stochastic system. \Sec{Sec:3} is devoted to the MSSM potential given in Eq.\eqref{MSSM potential}, where we investigate the behavior of the flat direction using an appropriate numerical simulation. In \Sec{Sec:4}, we examine the effect of non-Markovianity by applying the established numerical framework to simple models, namely $V=\lambda\phi^4,V=\mu\phi\chi+\lambda\phi^4$. Our main results are summarized in \Sec{Sec:5}. 

\section{Brief review of the stochastic formalism}
\label{Sec:2}

In this section, we briefly review the stochastic formalism, paying particular attention to its detailed structure, especially the origin of non-Markovianity, which is a distinctive feature of the stochastic formalism that differs from that arising in conventional open-systems approaches. 

Throughout this paper, let us fix the background spacetime to be de Sitter, neglecting the backreaction of perturbations on spacetime: 
\begin{align}
    \dd s^2 = - \dd t + a^2\delta_{ij}\dd x^i\dd x^j\quad \mathrm{with\ \ } a\propto e^{Ht}.
\end{align}
We consider a system of multiple scalar fields $\phi^I$, whose action is 
\begin{align}
    S=\int \dd^4 x \sqrt{-g}\com{-\frac{1}{2}g^{\mu\nu}\partial_\mu\phi^I\partial_\nu\phi^I - V(\phi^I)}
\end{align}
where $g^{\mu\nu}$ denotes the (inverse) spacetime metric, $V(\phi^I)$ is the scalar potential, and $I$ labels the field indices. For later convenience, we adopt the number of \efolds~$N$, defined via $\dd N = H\dd t$, as the time variable in what follows.

\subsection{Formulation}

In the stochastic formalism, fields are decomposed into IR and UV modes by introducing a coarse-graining scale or the cutoff scale 
\begin{align}
    k_\sigma(N) = \sigma a H
\end{align}
with $\sigma\ll 1$~\cite{Grain:2017dqa,Andersen:2021lii}. The separation is implemented by means of a window function $W(x)$, which, at leading order, is taken to be a step function $W(x)\rightarrow\theta(1-x)$. For a generic quantity $Q$, the IR and UV components are defined as
\begin{align}
    Q_\mathrm{IR} &\equiv \int\frac{\dd^3 k}{(2\pi)^3} e^{i \mathbf{k}\cdot\mathbf{x}} W\left(\frac{k}{k_\sigma(N)}\right) Q(N,\mathbf k) \equiv \int\frac{\dd^3 k}{(2\pi)^3}e^{i \mathbf{k}\cdot\mathbf{x}} Q_\mathrm{IR}(N,\mathbf k),\\
    Q_\mathrm{UV} &\equiv \int\frac{\dd^3 k}{(2\pi)^3} e^{i \mathbf{k}\cdot\mathbf{x}} \com{1-W\left(\frac{k}{k_\sigma(N)}\right)} Q(N,\mathbf k) \equiv \int\frac{\dd^3 k}{(2\pi)^3}e^{i \mathbf{k}\cdot\mathbf{x}} Q_\mathrm{UV}(N,\mathbf k).
\end{align}

Intuitively, the stochastic formalism proceeds as follows. The UV modes are treated quantum mechanically, typically within perturbative theory and assuming the Bunch-Davies vacuum. In contrast, the IR dynamics are effectively classical, and spatial gradient interactions can be neglected. As a result, each coarse-grained patch evolves independently in the separate-universe approximation~\cite{Sasaki:1998ug,Wands:2000dp,Lyth:2003im,Lyth:2004gb}. As UV modes cross the coarse-graining scale $k_\sigma(N)$, they are transferred into the IR sector, inducing stochastic fluctuations that drive the evolution of the IR background away from the homogeneous background. This continuous inflow of modes is modeled by stochastic noise terms, leading to Langevin equations as equations of motion for IR modes. 

Indeed, this picture can be confirmed and rigorously justified within the path integral formalism for open systems~\cite{Morikawa:1989xz,PerreaultLevasseur:2013kfq,Moss:2016uix, Pinol:2020cdp, Panda:2025tpu}, while alternative derivations using wave-functionals or density matrices also exist~\cite{Burgess:2014eoa,Burgess:2015ajz,Moss:2016uix,Gorbenko:2019rza,Cohen:2020php,Cohen:2021fzf,Cespedes:2023aal,Calzetta:2008iqa}. Starting from the action of the full system, it is separated into an only IR part, which constitutes the system of interest, and an UV-IR part, which represents an environment and interactions between the system and the environment. Integrating out the environmental degrees of freedom in the latter part yields the so-called influence action, which generically contains both real and imaginary parts\footnote{However, in the stochastic formalism, the influence action is purely imaginary. This property arises because, when integrating out the UV modes, the appropriate boundary conditions have to be specified~\cite{Tokuda:2017fdh,Tokuda:2018eqs}. This feature is specific to stochastic inflation and originates from the fact that the same field degrees of freedom are split into the system (IR modes) and the environment (UV modes) by the introduction of the cutoff, with the mode transfer from the UV sector to  the IR sector being unidirectional.}. Indeed, the effective dynamics of the system in the IR sector is no longer unitary. By introducing auxiliary fields (interpreted as stochastic noise sources) via Hubbard-Stratonovich transformation of the influence action, one obtains an effective action for the IR modes. This procedure leads to Langevin equations as equations of motion for classical fields\footnote{In the Schwinger-Keldysh formalism, the classical fields here are defined as the sum-average of the field configurations on the forward (usually denoted as ``$+$") and backward (usually denoted as ``$-$") time contours. These fields are also called retarded fields.}, given by
\begin{align}
    \dif{\phi_\mathrm{IR}^I}{N} &=\frac{\pi_{\mathrm{IR}}^I}{H} + \Xi_{\phi^I} \label{phi IR},\\
    \dif{\pi_\mathrm{IR}^I}{N} &=-3\pi_{\mathrm{IR}}^I - \frac{V_{,\phi^I}(\phi_{\mathrm{IR}})}{H} + \Xi_{\pi^I} \label{pi IR}
\end{align}
where $\pi^{I} \, (=\dot{\phi}^I)$ is the conjugate momentum of $\phi^I$, and $V_{,\phi^I}(\phi_{\mathrm{IR}})$ denotes the derivative of the potential with respect to  $\phi_\uIR^I$. The auxiliary fields $\Xi_{\phi^I},\Xi_{\pi^I}$ act as statistical noise terms. At leading order, these noise terms follow Gaussian statistics, with their variance\footnote{For simplicity, we focus on the dynamics of a single coarse-grained patch by invoking the separate-universe approximation. Consequently, the sinc function associated with correlations between patches at different spatial locations is neglected in the noise variance. }  given as
\begin{align}
    \Braket{\Xi_{X^I}(N)\Xi_{Y^J}(N')} &= \mathrm{Re}\mathcal{P}_{X^IY^J}(N,k_\sigma(N))\delta(N-N')\notag\\
    &={\left.\prn{\frac{k^3}{2\pi^2} \mathrm{Re} (X^{I}_\mathrm{UV}Y^{J*}_\mathrm{UV})}\right|_{k=k_\sigma(N)}} \delta_{IJ}\delta(N-N'). \label{variance of noise}
\end{align}
where $X,Y\in\{\phi,\pi\}$, and $I,J$ label the field species. Equivalently, one may introduce normalized noise variables $\xi_{X^I}$, which are white noises due to the use of a step function as a window function:
\begin{align}
    \Xi_{X^I}(N) = \sqrt{\mathcal{P}_{X^IX^I}(N,k_\sigma(N))}\xi_{X^I}(N)\quad \mathrm{with}\ \Braket{\xi_{X^I}(N)\xi_{X^I}(N')}=\delta(N-N'). \label{normalized noises}
\end{align}
As emphasized above, UV modes affect the IR dynamics when they cross the coarse-graining scale $k_\sigma(N)$. Note that to compute the noise terms, the values of UV modes at the coarse-graining scale are necessary, which can be obtained from UV modes evolving from deep inside the sub-horizon to the coarse-graining scale by their equations of motion on top of the evolving IR background:
\begin{align}
     0&= - \pdif{\phi_\mathrm{UV}^I(N,k)}{N}  +  \frac{\pi_\mathrm{UV}^I(N,k)}{H}  \label{phi UV}\\
     0&= - \pdif{\pi_\mathrm{UV}^I(N,k)}{N} -3\pi_\mathrm{UV}^I(N,k) - \frac{k^2}{a^2H}\phi_\mathrm{UV}^I(N,k) -  \frac{V_{,\phi^I\phi^J}(\phi_\mathrm{IR})}{H}  \phi_{\mathrm{UV}}^J(N,k)   \label{pi UV},
\end{align}
supplemented with the Bunch-Davies vacuum conditions in the deep sub-horizon.

\subsection{The method to solve the system}
\label{sec:post:inflation}

Let us comment on the structure of the equations. Regarding the UV modes, the equations for their mode functions \eqref{phi UV} and \eqref{pi UV} are formally similar to the linear equations of motion in cosmological perturbation theory by construction, except that the fixed, deterministic background field is replaced by the stochastic IR background. Hence, these equations cannot, in general, be solved analytically, and their power spectra cannot be calculated in a analytical form. Therefore, in order to determine the noise terms at each time step, the UV mode equations \eqref{phi UV} and \eqref{pi UV} have to be solved on top of the IR modes evolving by \eqref{phi IR} and \eqref{pi IR} at every step. Moreover, because the solution of Eqs. \eqref{phi UV} and \eqref{pi UV} depends on the entire history of the IR background, the resulting IR dynamics is inherently a non-Markovian system. In fact, when evaluating the noise terms needed to update the IR variable to the next time step, information about the IR quantities at slightly different times is required since the differential equations in time for the UV modes \eqref{phi UV} and \eqref{pi UV} must be solved. This explicitly implies that, in the Langevin equations \eqref{phi IR} and \eqref{pi IR}, knowledge of the IR quantities at a single instant is insufficient to determine their future evolution to the next time step. By definition, this places the system beyond the class of Markovian stochastic processes. Consequently, a fully consistent treatment, including non-Markovianity, requires numerical simulations in general, which are computationally heavy, as the UV modes equations must be solved for all relevant modes at each time step. The details of the numerical algorithm are presented in \App{Sec:A}. 

However, if the characteristic time scale over which the UV modes evolve as an environment is much shorter than that of the IR modes as a system, then the IR modes do not change significantly in time compared to the evolution of the UV modes. Therefore, the UV dynamics at a given time is effectively influenced only by the instantaneous values of the IR modes. Moreover, the noise terms appearing in the Langevin equations for IR modes \eqref{phi IR} and \eqref{pi IR} can be determined solely by the IR modes only at that instantaneous time, implying the absence of memory effects and thus leading to Markovian systems~\cite{Burgess:2014eoa,Burgess:2015ajz,Mirbabayi:2020vyt}. Indeed, in this regime, the IR modes can be treated as approximately constant within each time step when solving the UV equations \eqref{phi UV} and \eqref{pi UV}. Consequently, these equations could become solvable on time scales where IR modes may be regarded as fixed constants. In particular, for scalar fields with constant effective masses in de Sitter spacetime, the UV mode functions are known analytically with the Bunch-Davies vacuum. In these analytical solutions, at every time step, IR quantities at this time should be substituted into the part of the constant effective masses. In the case where the subscript of the Hankel function, $\nu$, is a real number, these UV mode solutions evaluated at the coarse-graining scale $k=k_\sigma(N)$ are given by
\begin{align}
    {\phi^I_\mathrm{UV}}(N, k_\sigma(N)) &= \sqrt{\frac{\pi}{4k_\sigma^3}} H  e^{i\frac{\pi\nu}{2}+i\frac{\pi}{4}}\sigma^{3/2} H^{(1)}_{\nu}(\sigma) ,\label{Markovphi}\\
    {\pi^I_\mathrm{UV}}(N, k_\sigma(N)) &= \sqrt{\frac{\pi}{4k_\sigma^3}} H  e^{i\frac{\pi\nu}{2}+i\frac{\pi}{4}}\sigma^{3/2} \com{\prn{\nu - \frac{3}{2}}H^{(1)}_{\nu}(\sigma) -\sigma H^{(1)}_{\nu-1}(\sigma)   }\label{Markovpi}
\end{align}
where 
\begin{align}
    \nu \equiv \sqrt{\frac{9}{4} - \frac{V_{,\phi^I\phi^I}(\phi_\mathrm{IR})}{H^2}}.    
\end{align} 
Here, for simplicity, we neglect the non-diagonal components of the mass matrix $V_{\phi^I\phi^J}$, that is, the second order derivative in terms of scalar fields. When such terms are present, one should work in a basis that diagonalizes this matrix. In fact, it is always possible to do that due to the linearity of the UV modes equations. 

There exists another practical computational strategy, known as the recursive approach~\cite{PerreaultLevasseur:2013eno,PerreaultLevasseur:2013kfq,PerreaultLevasseur:2014ziv,Figueroa:2021zah}. This method iteratively solves the UV mode equations using progressively refined statistical information about the IR background, and it is thereby expected to converge toward the true stochastic dynamics. Concretely, one begins by solving the UV mode equations on a homogeneous background, rather than on the IR background, so that the background is completely deterministic. In this initial step, the mass matrix is approximated as $V_{\phi^I\phi^J}(\phi_\uIR)\rightarrow {V_{\phi^I\phi^J}(\bar \phi)}$, and Langevin equations are solved using the noise terms computed from these UV mode solutions. In the subsequent interaction, the UV mode equations are solved again, this time with the background replaced by the statistical IR configuration obtained from the previous step, such as $V_{\phi^I\phi^J}(\bar \phi)\rightarrow \Braket{V_{\phi^I\phi^J}(\phi_\uIR)}$. The corresponding Langevin equations are then solved using the updated noise terms. This procedure can be continued iteratively. For example, when the initial conditions of IR modes are taken to be at the origin, assumed throughout this paper, the first iteration can be solved analytically. In this case, one recovers the Markovian solutions \eqref{Markovphi} and \eqref{Markovpi} with the effective masses vanishing. Therefore, the noise terms at the first iteration reduce to the familiar expression $H/(2\pi) \xi$.

In the following sections, we numerically solve the Langevin equations using the numerical framework described in \App{Sec:A}, in which the UV mode equations are solved simultaneously for all modes at each time step. The coarse-graining parameter (or the cutoff) is taken to be $\sigma=0.01$. Random numbers are generated using the Mersenne Twister method~\cite{Mersenne} throughout this paper, and normalized Gaussian noise variables $\xi_{X^I}$ are constructed via the Box-Muller method with the random numbers. For practical convenience, we work with the following dimensionless variables:
\begin{align}
    &\phi^I_\uIR \rightarrow \widetilde{\phi}^I\equiv\frac{\phi_\uIR^I}{H},\quad \pi^I \rightarrow \widetilde{\pi}^I\equiv\frac{\pi^I_\uIR}{H^2},\\
     &\mathcal{P}_{\phi^I\phi^I} \rightarrow \widetilde{\mathsf N}_{\phi^I}^2\equiv\frac{\mathcal{P}_{\phi^I\phi^I}}{H^2},\quad \mathcal{P}_{\pi^I\pi^I} \rightarrow \widetilde{\mathsf N}_{\pi^I}^2\equiv\frac{\mathcal{P}_{\pi^I\pi^I}}{H^4}.
\end{align}
In \Sec{Sec:3}, we apply this numerical framework to the MSSM potential and discuss its phenomenological consequences, comparing our results with those of previous works. \Sec{Sec:4} is devoted to elucidating the differences between the full non-Markov computation employed here and commonly used approximations, such as Markovian treatment, through illustrative examples.  
%

\section{Practice 1: Numerical analysis for flat and non-flat direction systems}\label{Sec:3}

In this section, we consider a system of flat directions coupled to non-flat directions in scalar field space and examine their evolutions within the stochastic formalism. As a concrete model, let us consider supersymmetric standard models, which are known to possess a large number of flat directions. In particular, in order to facilitate a direct comparison with previous studies~\cite{Enqvist:2011pt,Kawasaki:2012bk}, we analyze the same model whose scalar potential is given by 
\begin{align}
    V = \frac{1}{2}\lambda_e^2 (\phi^2 + h^2)\bar e^2 + \frac{1}{8}g_2^2h^4 + \frac{1}{8}g_1^2(h^4 + 4\bar e^4 - 4h^2 \bar e^2) + \frac{\phi^6}{\Mp^2}. \label{MSSM potential}
\end{align}
Here, $\lambda_e$, $g_1$, and $g_2$ denote coupling constants. This potential is motivated by the $LH_u$ flat direction in MSSM. The $\phi$ corresponds to a flat direction, while $\bar e$ and $h$ denote non-flat directions coupled to the flat direction. (see Sec.2 of \cite{Enqvist:2011pt} for details). The final term represents a non-renormalizable term suppressed by the Planck mass $\mpl\sim 10^{19}\mathrm{GeV}$; in the numerical simulation, we neglect this contribution since we restrict our attention to the regime in which it is subdominant. In the absence of tree-level Hubble-induced masses\footnote{Within supergravity framework, flat directions generically acquire effective masses of order the Hubble scale during inflation~\cite{PhysRevD.51.6847,DINE1996291,PhysRevLett.75.398}. However, such contribution can be neglected, for example, in the D-term inflation scenario~\cite{PhysRevD.51.6847} or by imposing a Heisenberg symmetry on K\"{a}hler potential~\cite{PhysRevLett.75.398}.}, the time evolution of a flat direction without stochastic kicks, is governed by one-loop radiative corrections or non-renormalizable terms. Once stochastic noise sourced by UV quantum corrections is included, however, the dynamics could be modified. Refs.~\cite{Enqvist:2011pt,Kawasaki:2012bk} have investigated such interacting systems consisting of flat and non-flat directions within the stochastic framework. 

Let us briefly summarize their analysis methods and main claims. In Ref.~\cite{Enqvist:2011pt}, they solve the Langevin equations for the system while neglecting the effective masses arising from the IR potential in the noise sector, \ie\ they adopt $H/(2\pi)\xi$ for massless free fields. They then conclude that the variance of the flat direction saturates and approaches a stationary state, arguing that the non-flat directions hinder the evolution of the flat directions through the couplings between them. However, in this approach, although the contributions of the effective mass to the drift terms are taken into account, those to the noise terms are ignored, which is inconsistent with the construction of the stochastic formalism reviewed in \Sec{Sec:2} unless all fields are free. In contrast, Ref.~\cite{Kawasaki:2012bk} attempts to incorporate the effects of effective masses into the noise terms of the Langevin equations and claims that the flat direction finally becomes flat and its evolution is determined by one-loop corrections or non-renormalizable terms as usual. Nevertheless, the methodology employed there is neither complete nor fully consistent. In particular, the effective mass matrix is not diagonalized, and additional ad hoc subtractions are introduced without a clear justification, which they refer to as the contributions from zero-point fluctuations (corresponding to Sec.~2.2 of Ref.~\cite{Kawasaki:2012bk}). In addition, in both Refs.~\cite{Enqvist:2011pt,Kawasaki:2012bk}, the cross correlations of noise terms among fields and their momenta are also absent, even though the omission is expected to have a limited quantitative impact due to the smallness of the relevant momenta. 

In this paper, we directly compute the Langevin equations \eqref{phi IR} and \eqref{pi IR} by solving the UV mode equations \eqref{phi UV} and \eqref{pi UV} for all relevant modes at every time step. This procedure correctly incorporates all contributions from effective masses into the noises. For the potential \eqref{MSSM potential}, the effective mass matrix appearing in the UV mode equation \eqref{pi UV} is
\begin{align}
    \frac{V_{\phi^I\phi^J}}{H^2} = \com{\begin{array}{ccc}
        \lambda_e^2 \widetilde{\bar e}^2 & 2\lambda_e^2\widetilde{\phi}\widetilde{\bar e} & 0\\
        2\lambda_e^2\widetilde{\phi}\widetilde{\bar e} & \lambda_e^2\widetilde{\phi}^2 + (\lambda_e^2 - g_1^2)\widetilde{h}^2 + 6g_1^2\widetilde{\bar e}^2 & 2(\lambda_e^2 - g_1^2)\widetilde{\bar e} \widetilde{h}\\
        0 & 2(\lambda_e^2 - g_1^2)\widetilde{\bar e}\widetilde{ h} & (\lambda_e^2 - g_1^2)\widetilde{\bar e}^2 + \frac{3}{2}(g_1^2 + g_2^2)\widetilde{h}^2 
    \end{array}}, \label{effective mass matrix}
\end{align}
which contains non-diagonal components\footnote{If one adopts the Markovian approximation using the analytical solutions such as \eqref{Markovphi} and \eqref{Markovpi}, this effective mass matrix must first be diagonalized. This is because the UV mode functions admit these analytical Markovian solutions only in the new basis of fields that diagonalize this matrix. It should be emphasized, however, that the BD vacuum must be imposed as the initial condition for the original field variables, rather than for the new basis defined above.}. Focusing on the two field subsystem ($\phi,\bar e$), which is the case studied below, the dimensionless fields $\widetilde{\phi},\widetilde{\bar e}$ are fully mixed in the corresponding $2\times 2$ submatrix of the effective mass matrix \eqref{effective mass matrix}.

\subsection{Setup}

To allow a direct comparison with the previous studies \cite{Enqvist:2011pt,Kawasaki:2012bk}, we perform numerical simulations using the same setup. Specifically, we initialize all fields and their momenta at the origin at $N=0$:
\begin{align}
    \widetilde{\phi} (0) = \widetilde{\pi_{\phi}}(0) = 0,\ \widetilde{\bar e} (0) = \widetilde{\pi_{\bar e}}(0) = 0,\ \widetilde{h} (0) = \widetilde{\pi_{ h}}(0) = 0.
\end{align}
The coupling constants in the potential \eqref{MSSM potential} are chosen as $\lambda_e =g_1=g_2=1.0 $\footnote{Ref.~\cite{Enqvist:2011pt} shows that the strength of couplings primarily affects the time scale of the dynamics. Therefore, to understand the asymptotic evolution of the flat direction, it suffices to examine the strong coupling case, in which the dynamics proceeds more rapidly. We note, however, that the formulation for the stochastic formalism involves a subtlety in the presence of strong couplings. These issues lie beyond the scope of this paper and will be mentioned in \Sec{Sec:5}.\label{footnote: couplings}}. With this choice, the Higgs field decouples from the other fields, and we therefore focus on the dynamics of the two fields $\phi,\bar e$ in what follows. The number of stochastic realizations is taken to be $3 \times10^4$, and the time step $\dd N$ is set to $0.1$. Although this step is not sufficiently small to obtain high-precision results (the numerical error is at the level of $\mathcal O(1)\%$), it is still adequate for capturing the qualitative behavior and order estimations of the fields. 

\subsection{Results and discussion}

\begin{figure}[h]
 \begin{center} 
  \subfigure{
   \includegraphics[width=.5\columnwidth]{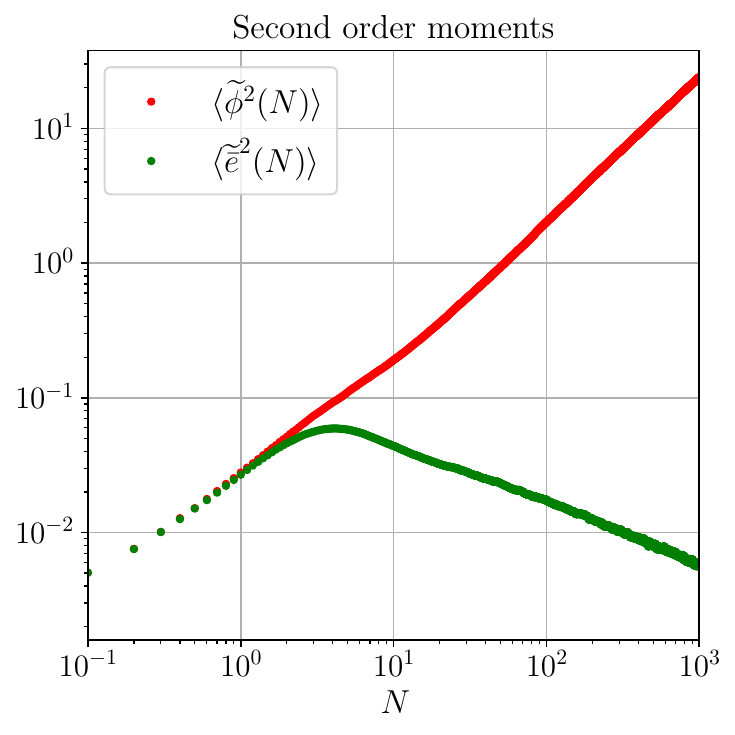}
  }~
  \subfigure{
   \includegraphics[width=.5\columnwidth]{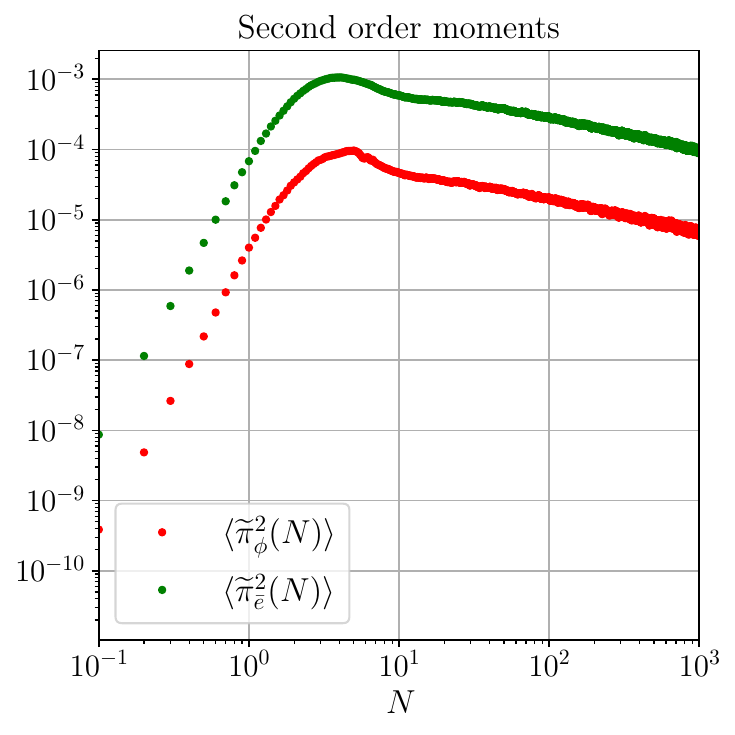}
  }\\
  \subfigure{
   \includegraphics[width=.5\columnwidth]{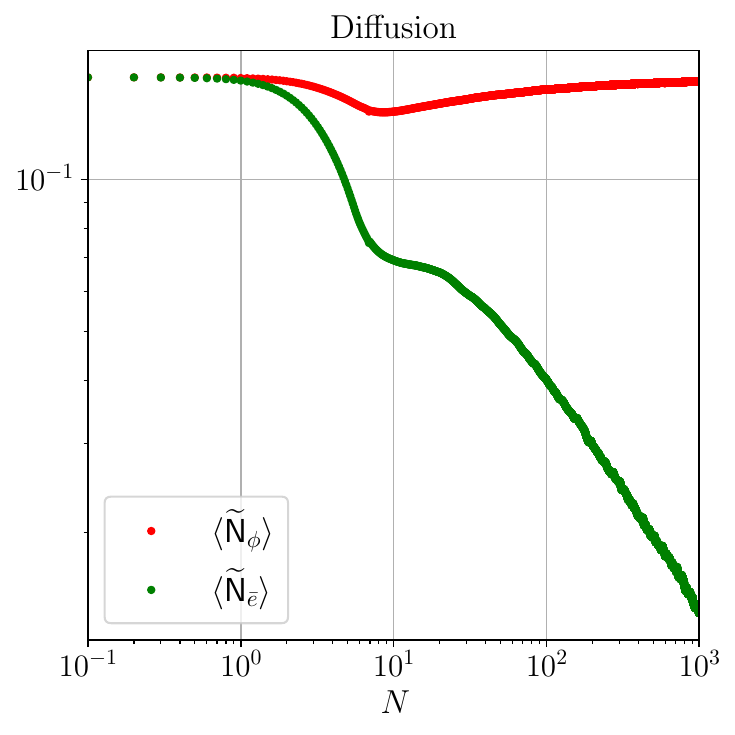}
  }~
  \subfigure{
   \includegraphics[width=.5\columnwidth]{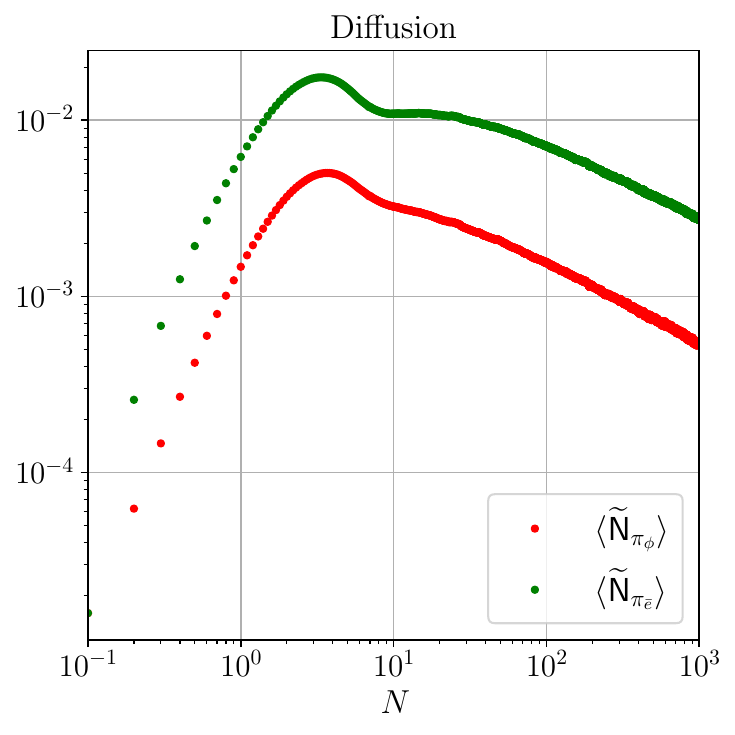}
  }
  \caption{Second order moments (upper panels) and the amplitude of diffusion terms \eqref{normalized noises} in Langevin equations (lower panels) for IR fields (left panels) and their momenta (right panels) in the MSSM system with the potential \eqref{MSSM potential}. 
  } 
  \label{Fig:MSSM}
 \end{center}
\end{figure}

The results of the numerical simulation are displayed in \Fig{Fig:MSSM}. From these results, we conclude that the flat direction eventually remains flat, which apparently agrees with Ref.~\cite{Kawasaki:2012bk}. It can be understood as follows. The interactions between the flat direction $\phi$ and the non-flat direction $\bar e$ generate effective masses for the UV fluctuations on top of the IR backgrounds, contributing to the noise terms in the Langevin equations. In the potential \eqref{MSSM potential}, $\bar e$ has a self-coupling term whereas $\phi$ does not, and then the effective mass of $\bar e$ receives a larger contribution than that of $\phi$. At early times (until $N\sim 1$), the diffusion strengths of $\phi$ and  $\bar e$ characterized by $\widetilde{\mathsf{N}}_{\phi}$ and $\widetilde{\mathsf{N}}_{\bar e}$ are identical, and hence the second order moments of $\phi$ and $\bar e$ initially grow at the same rate. However, as the contributions from the effective masses induced by a self-coupling become dominant, their diffusion rates begin to differ, as shown in the left panels of \Fig{Fig:MSSM}. After this splitting occurs, our numerical simulations show that the variance of $\bar e$ decreases, while that of the flat direction $\phi$ continues to grow. Even though the variance of $\bar e$ becomes smaller, $\bar e$ effectively becomes more massive (the amplitude of its noise $\widetilde{\mathsf{N}}_{\bar e}$ is suppressed) because the growing flat direction $\phi$ induces a large effective mass to $\bar e$. In contrast, $\phi$ eventually behaves as an exactly flat direction, since the contributions to the effective masses become negligible again due to the sufficient suppression of the variance of $\bar e$. Indeed, we also confirm that the PDFs for $\phi,\bar e$ show that the flat direction $\phi$ continues to diffuse and remain flat, whereas the non-flat direction $\bar e$ stays localized near the origin despite undergoing an initial diffusion phase (until $N\sim3$). From Fig.~\ref{Fig:MSSM}, one can see that the momenta remain statistically subdominant throughout the evolution, so their effect on the asymptotic behavior of IR modes is negligible.

In conclusion, the flat direction eventually becomes flat, and its late-time evolution is governed by one-loop corrections or non-renormalizable terms. This conclusion can be quantitatively confirmed as follows. In order to discuss the dynamics of the flat direction $\phi$, let us focus on the effect of the potential $V_{,\phi}$ appearing in its dynamical equation, \ie\ Langevin equation. For instance, the non-renormalizable term in \eqref{MSSM potential} gives rise to $V_\phi/H^3\sim \widetilde\phi^5(H/\mpl)^2$ while the coupling between $\phi$ and $\bar e$ contributes to the potential drift as $V_\phi/H^3\sim \lambda_e^2\widetilde\phi\widetilde{\bar e}^2$. In the absence of the non-renormalizable term, $\phi$ remains an exactly flat direction until $V_\phi/H^3\sim \lambda_e^2\widetilde\phi\widetilde{\bar e}^2$ becomes of order unity. In other words, $\phi$ can grow freely until $\widetilde{\phi}\gtrsim 10^4\lambda_e^{-2}$ because $\widetilde{\bar e}\lesssim 10^{-2}$ at least from the result shown in \Fig{Fig:MSSM}. On the other hand, the drift induced by the non-renormalizable term $V_\phi/H^3\sim \widetilde\phi^5(H/\mpl)^2$ reaches order unity faster in the case of $H/\mpl\gtrsim 10^{-10}\lambda_e^{5}$, which is satisfied except for very low-energy inflation scenarios. 

Let us emphasize that although our discussion and conclusion are the same as those of Ref.~\cite{Kawasaki:2012bk}, our numerical simulations correctly capture the effective masses by construction, while Ref.~\cite{Kawasaki:2012bk} does not properly consider them, in particular, the diagonalization of the effective mass matrix \eqref{effective mass matrix}. Moreover, they report that the saturation of the flat direction occurs unless an artificial subtraction from the full results is performed; however, our result shows the above conclusion without such ambiguity.

\section{Practice 2: Identifying non-Markovian contributions}\label{Sec:4}

As discussed in \Sec{Sec:2}, the full numerical treatment solves the Langevin equations \eqref{phi IR} and \eqref{pi IR} with noises determined by solving UV mode equations \eqref{phi UV} and \eqref{pi UV} at each time step, which requires knowledge of the entire past history of the IR system. This feature is referred to as non-Markovianity. In this section, we investigate the impact of such non-Markovian effects by comparing the full numerical results including non-Markovianity with those under the Markovian approximation \eqref{Markovphi} and \eqref{Markovpi}, as well as with the commonly used massless free field noise prescription such as $H/(2\pi)\xi$ for fields.

\subsection{Setup}

In the following, let us consider two simple models:
\begin{description}
    \item[Single field $V={\lambda}\phi^4$]\mbox{} \\
        The model describes a massless self-interacting scalar field. This potential is bounded from below and provides a barrier for large field values. Therefore, the system eventually approaches a stationary state in which the effect of the potential drift is balanced by that of diffusion due to noise. We numerically solve the following Langevin equations together with the equations of UV modes that determine their noises \eqref{phi UV} and \eqref{pi UV}:
        \begin{align}
        \dif{\widetilde{\phi}}{N} &=\widetilde{\pi} + \widetilde{\mathsf N}_\phi\xi_\phi \label{single phi},\\
        \dif{\widetilde{\pi}}{N} &=-3\widetilde{\pi} - {4\lambda \widetilde{\phi}^3} + \widetilde{\mathsf N}_\pi\xi_\pi \label{single pi}.
        \end{align}
        The initial conditions at $N=0$ are taken to be
        \begin{align}
            \widetilde{\phi} (0) = \widetilde{\pi}(0) = 0.\label{IC eg1}
        \end{align}

    \item[Multi-field $V= \mu^2 \phi\chi + \lambda \phi^4$]\mbox{} \\
        This model consists of two interacting scalar fields, one of which possesses a self-interaction. Even though the potential is apparently unstable, it is implicitly assumed that additional higher-order terms stabilize the potential at large field values by adding proper terms. In the present analysis, we focus on the regime in which such stabilizing terms are negligible. For simplicity, we neglect the momenta of the fields and focus solely on fields equations. We therefore numerically solve the following Langevin equations for the fields: 
    \begin{align}
    \dif{\widetilde{\phi}}{N} &= -\frac{\mu^2}{3H^2} \widetilde{\chi} - \frac{4}{3}\lambda \widetilde\phi^3 + \widetilde{\mathsf N}_\phi\xi_{\phi}, \label{multi phi}\\
    \dif{\widetilde{\chi}}{N} &= -\frac{\mu^2}{3H^2} \widetilde{\phi}  + \widetilde{\mathsf N}_\chi\xi_{\chi}\label{multi chi}.
    \end{align}
    Here again, noises are computed by solving the equations of UV modes \eqref{phi UV} and \eqref{pi UV}. The initial conditions at $N=0$ are also chosen as
        \begin{align}
            \widetilde{\phi} (0) = \widetilde{\chi}(0) = 0. \label{IC eg2}
        \end{align}
\end{description}

\subsection{Results and discussion}\label{Sec:42}

\subsubsection{Single field $V={\lambda}\phi^4$}

\begin{figure}[h]
 \begin{center} 
  \subfigure{
   \includegraphics[width=.5\columnwidth]{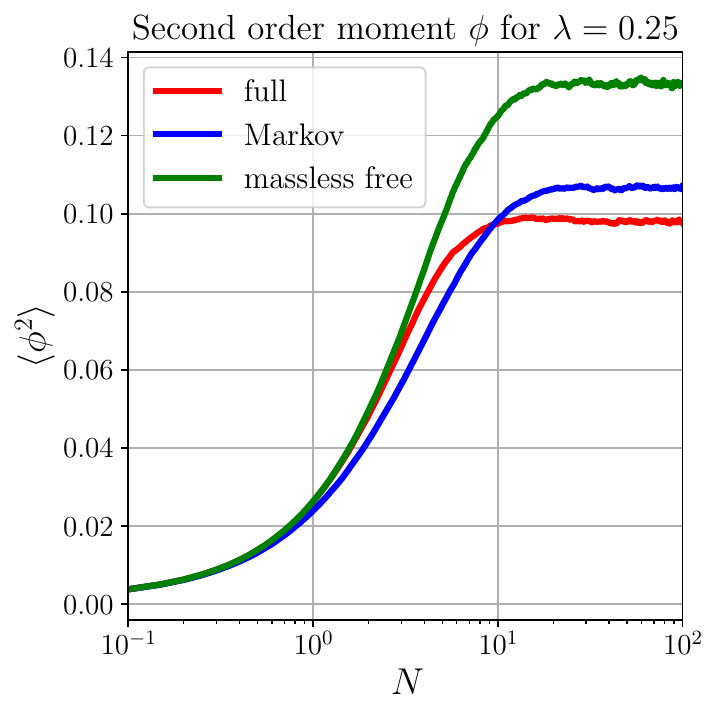}
  }~
  \subfigure{
   \includegraphics[width=.5\columnwidth]{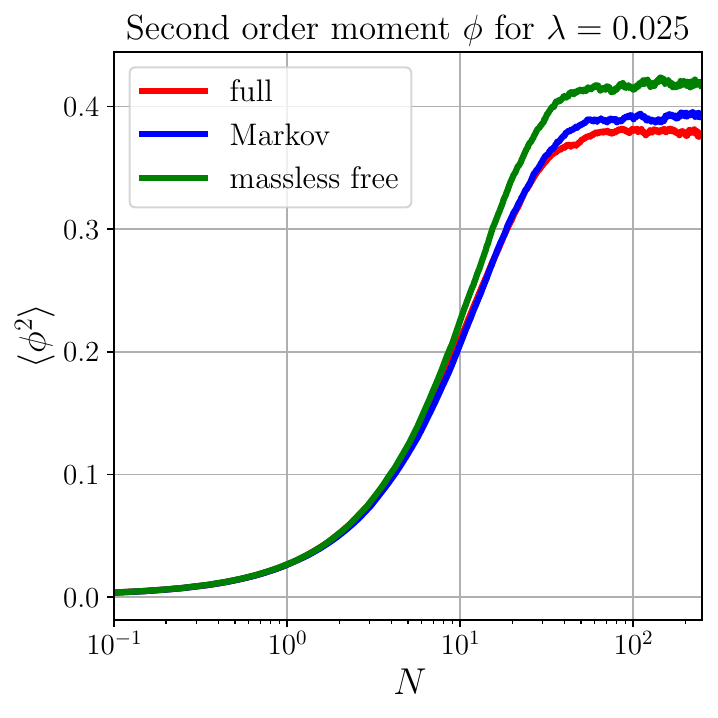}
  }\\
  \subfigure{
   \includegraphics[width=.5\columnwidth]{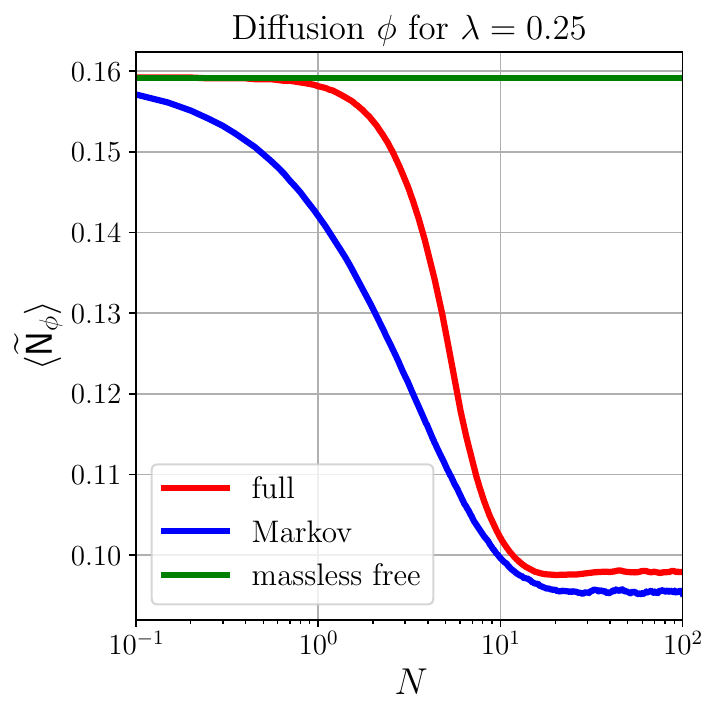}
  }~
  \subfigure{
   \includegraphics[width=.5\columnwidth]{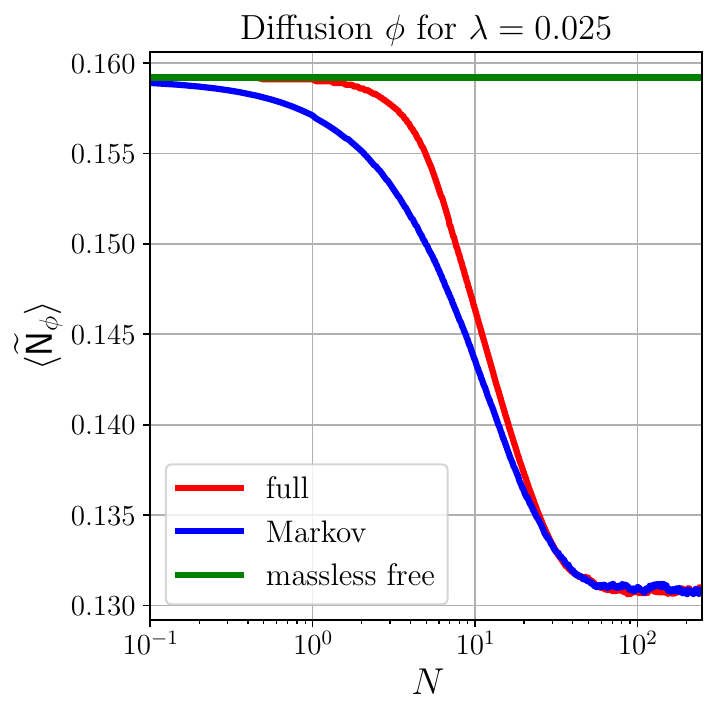}
  }
  \caption{Second order moments (upper panels) and the amplitude of diffusion terms \eqref{normalized noises} in Langevin equations \eqref{single phi} and \eqref{single pi} (lower panels) of the field for the potential $\lambda\phi^4$ with $\lambda=0.25$ (left panels) and $\lambda = 0.025$ (right panels). The number of stochastic realization are $ 1.0\times10^5$ and $5\times 10^4$ for the former and the latter cases. The time step is $\dd N=0.05$.} 
  \label{Fig:example11}
 \end{center}
\end{figure}

\begin{figure}[h]
 \begin{center} 
  \subfigure{
   \includegraphics[width=.5\columnwidth]{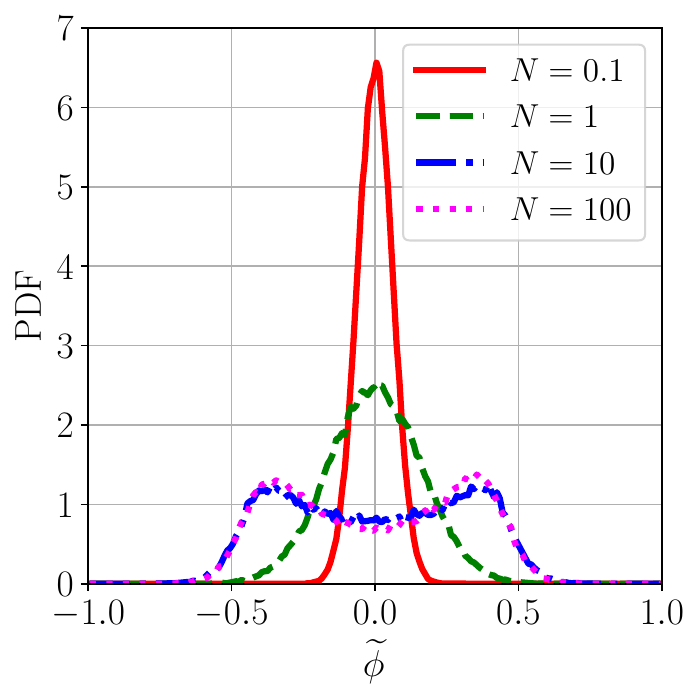}
  }~
  \subfigure{
   \includegraphics[width=.5\columnwidth]{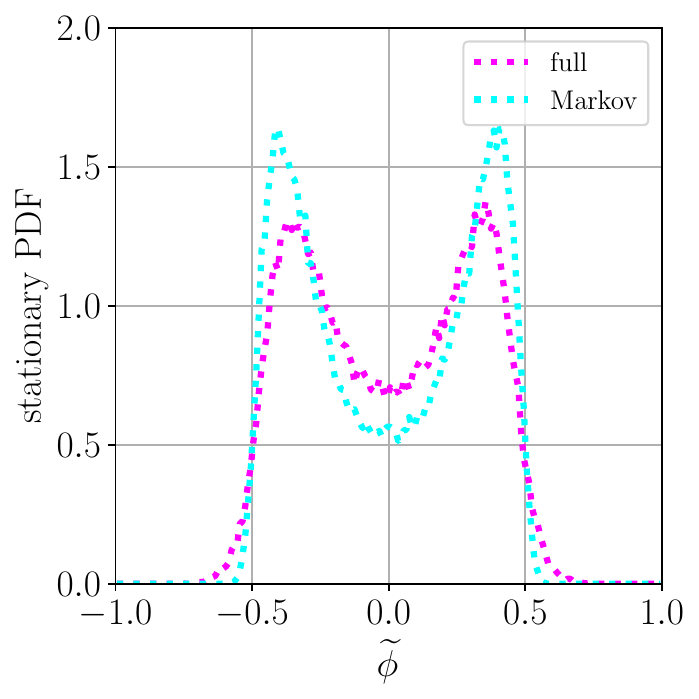}
  }
  \caption{\textit{left}: PDF of $\widetilde{\phi}$ for the potential $\lambda\phi^4$ with $\lambda=0.25$, obtained from the full non-Markov simulation (red) in the left panels of \Fig{Fig:example11}. \textit{right}: PDF for a stationary state at $N=100$ for $\lambda=0.25$ in both the non-Markov and the Markov cases.} 
  \label{Fig:example11PDF}
 \end{center}
\end{figure}

The results of these simulations are shown in \Figs{Fig:example11}, \ref{Fig:example11PDF} and \ref{Fig:example12}. Since the momenta remain smaller than the fields enough to be neglected as a first step, the following argument focuses on the field variables. The corresponding results for the momenta, shown in \Fig{Fig:example12} are therefore presented in \App{Sec:B}. In these figures, the red, blue, and green curves represent, respectively, the amplitude of noises computed in the full numerical simulation including non-Markovianity (by solving the equations of UV modes), those computed using Markovian approximated solutions \eqref{Markovphi} and \eqref{Markovpi}, and massless free field noises which correspond to the Markovian case with vanishing effective masses \ie\ $\widetilde{\mathsf N}_\phi = 1/(2\pi)$. Broadly speaking, in all cases $\phi$ initially undergoes stochastic diffusion driven by the noise and eventually approaches an equilibrium state due to the confining shape of the potential, as mentioned above. This asymptotic variance can be approximately estimated by the formula: $\Braket{\widetilde\phi^2}=\frac{3}{8\pi^2 V_{\phi\phi}/H^2}$  valid for $\phi \ll 1$. Nevertheless, these figures clearly demonstrate quantitative differences among the three cases: full non-Markovian noise (red), Markovian approximated noise (blue), and massless free field noises (green). In particular, the massless free field prescription fails to incorporate the effects of effective masses, as in the previous section, and hence leads to overestimation of the diffusion amplitudes (see the lower panels of \Figs{Fig:example11} and \ref{Fig:example12}). For this reason, in the following discussion, we particularly concentrate on the comparison between the fully non-Markovian and Markovian approximated cases.   

Let us now examine the memory effects associated with non-Markovianity in greater detail, focusing on the stronger coupling case $\lambda=0.25$ (left panels in \Fig{Fig:example11}), for which non-Markovianity effects are more pronounced. We highlight the following two points in particular. First, the diffusion amplitude associated with the Markovian noise decreases faster than that of the non-Markovian noise once $\phi$ acquires its effective mass. This behavior directly reflects a memory effect. This is because both the field and its momentum are close to zero at early times due to the initial conditions \eqref{IC eg1}, and then the non-Markovian noise retains information about the system's past evolution near the origin, whereas the Markovian noise depends only on the instantaneous values of the IR variables. More precisely, the Markovian noise, being determined by the IR modes only at a given time, tends to overestimate the contributions of the effective mass (or, equivalently, to underestimate the diffusion amplitude $\widetilde{\mathsf N}_\phi$) when $\phi$ and $\pi$ acquire relatively larger values. Therefore, in the fully non-Markovian calculation, the field grows more rapidly than in the Markovian approximation until the potential barrier becomes relevant. Once the potential drift plays a role, the field value in the non-Markovian case subsequently decreases more rapidly than in the Markovian case. Eventually, in both treatments, the system approaches a stationary state where the effects of noise and the potential barrier balance each other.

The other point worth mentioning concerns the stationary configurations reached after a sufficiently long evolution ($N\gtrsim30$). Using the noise evaluated in this stationary regime, we compute the PDF of $\phi$, shown in \Fig{Fig:example11PDF}. Both the non-Markovian and Markovian cases exhibit a bimodal structure, indicating that the most probable field values are localized around two peaks resulting from the balance between the potential barrier and stochastic diffusion. The right panel of \Fig{Fig:example11PDF}~shows that the Markovian case yields a more sharply localized distribution around the peaks, whereas the non-Markovian PDF is broader, extending toward both smaller and larger field values. This behavior can be understood as follows. In the equilibrium distribution, the absolute values of the field smaller than those of the peaks, located near the origin, are realized more frequently than larger values.\footnote{
In the case of $\lambda=0.25$, the typical magnitudes of $\phi$ and $\widetilde{\mathsf N}_\phi$ in the stationary state are both of the order of $0.1$, since $\Braket{\widetilde\phi^2}\sim\mathcal{O}(0.1)$ implies $\phi\sim\mathcal{O}(0.1)$. Consequently, even though the system is statistically stationary (so that the PDF does not evolve in time), the value of $\phi$ can change significantly at the next step. When $\phi$ evolves to a new value at the subsequent time step, transitions toward smaller field values are statistically favored over the transitions toward larger values due to the potential drift.} 
Around the peaks, the Markovian noise neglects the history in which $\phi$ spends more time at smaller values than at larger ones, by definition of the Markovian noise where the noise amplitude is evaluated solely based on the instantaneous field value at the peaks. Therefore, the Markovian noise statistically overestimates the contribution of the effective mass and thus underestimates the diffusion amplitude compared to the full Markovian case. This interpretation is consistent with the left bottom figure of \Fig{Fig:example11}. As a result, the Markovian dynamics leads to a more localized in the bimodal type PDF. For such a distribution, the second order moment in the non-Markovian case can be larger than that in the Markovian case, whilst we also confirm that higher order moments, specifically those of order six and above, are larger in the non-Markovian case, reflecting the broader tails of the distributions.

The two points discussed above illustrate the impact of the non-Markovian effects. In addition, by comparing the left and right panels of \Figs{Fig:example11}, we confirm that the influence of non-Markovianity diminishes as the coupling $\lambda$ decreases. As discussed in \Sec{Sec:2}, this behavior can be understood as follows. Within the stochastic formalism, the interaction between the system (IR modes) and the environment (UV modes) is governed by the strength of the self-couplings. As the couplings become weaker, the environment is less sensitive to variations in the system, or the characteristic timescale of the UV modes is much smaller than that of the IR modes. Consequently, the memory effects associated with non-Markovianity become less significant.

\subsubsection{Multifield $V= \mu^2 \phi\chi + \lambda \phi^4$}

\begin{figure}[h]
\centering

\begin{minipage}{0.32\columnwidth}
  \centering
  \includegraphics[width=\linewidth]{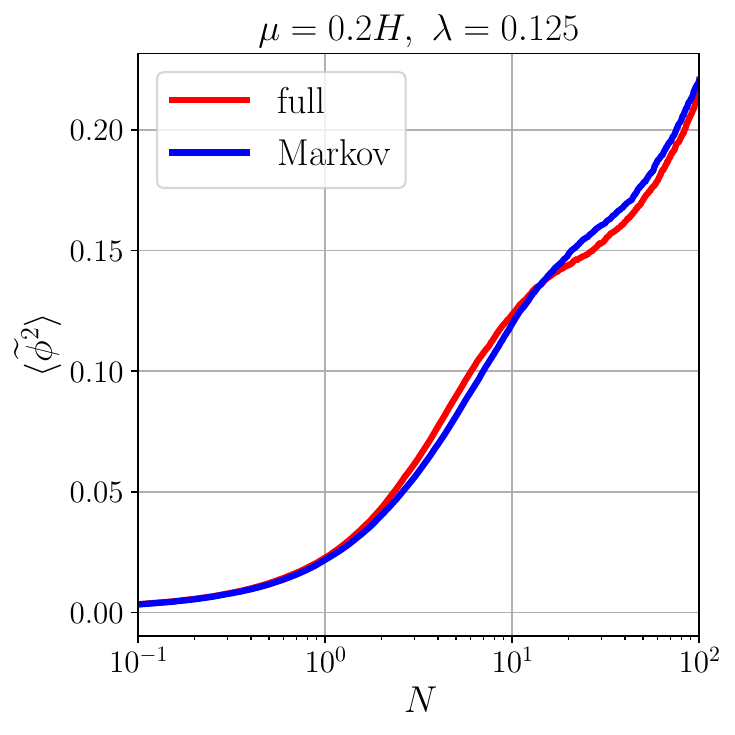}
\end{minipage}\hfill
\begin{minipage}{0.32\columnwidth}
  \centering
  \includegraphics[width=\linewidth]{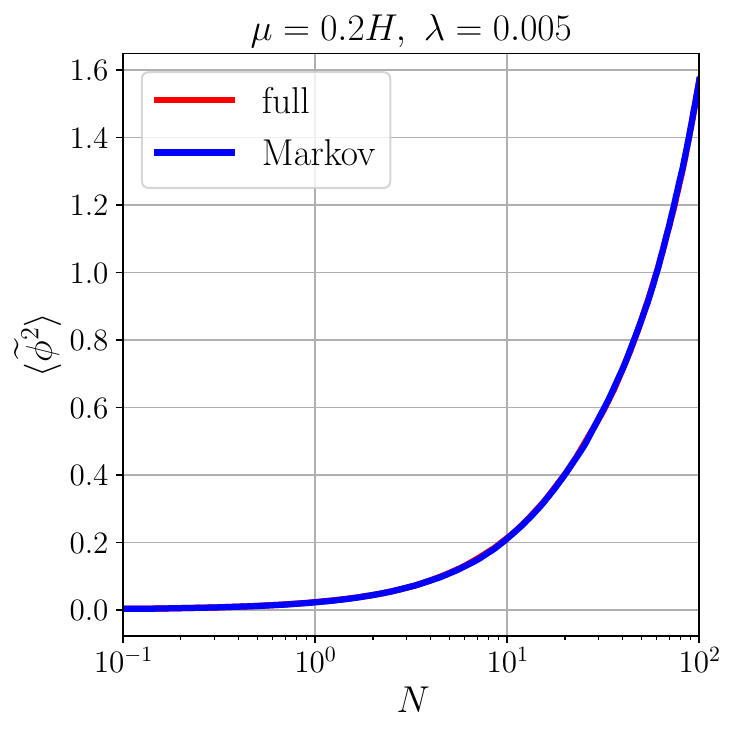}
\end{minipage}\hfill
\begin{minipage}{0.32\columnwidth}
  \centering
  \includegraphics[width=\linewidth]{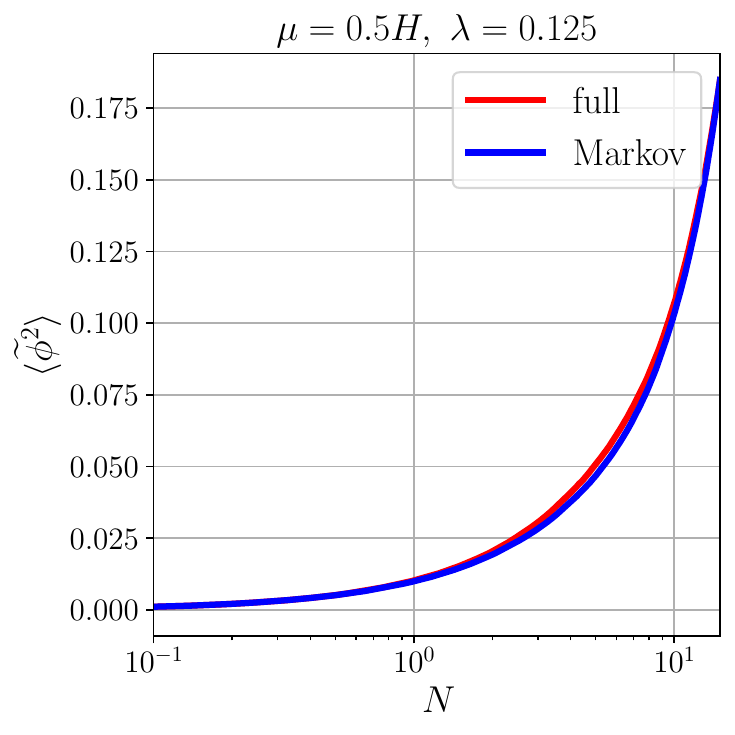}
\end{minipage}

\vspace{2mm}

\begin{minipage}{0.32\columnwidth}
  \centering
  \includegraphics[width=\linewidth]{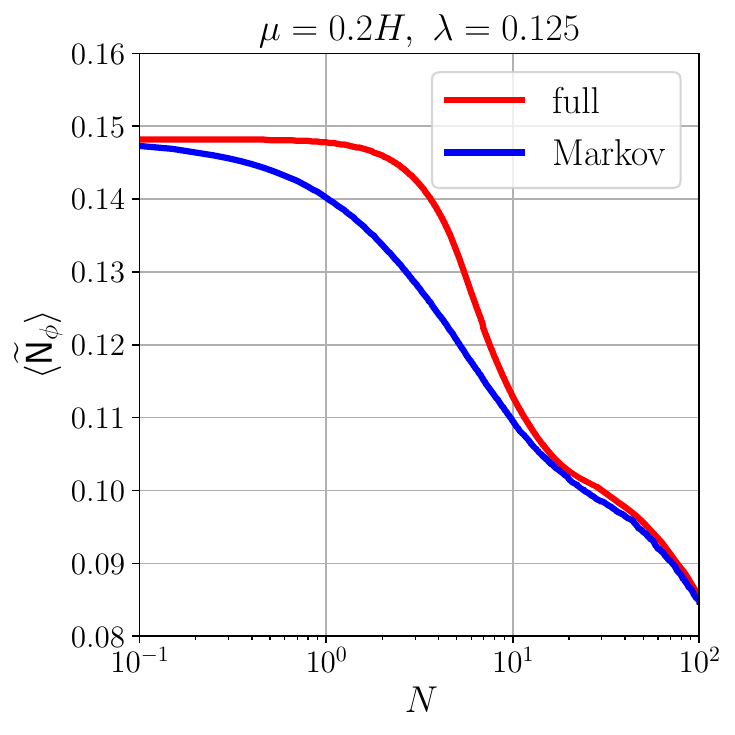}
\end{minipage}\hfill
\begin{minipage}{0.32\columnwidth}
  \centering
  \includegraphics[width=\linewidth]{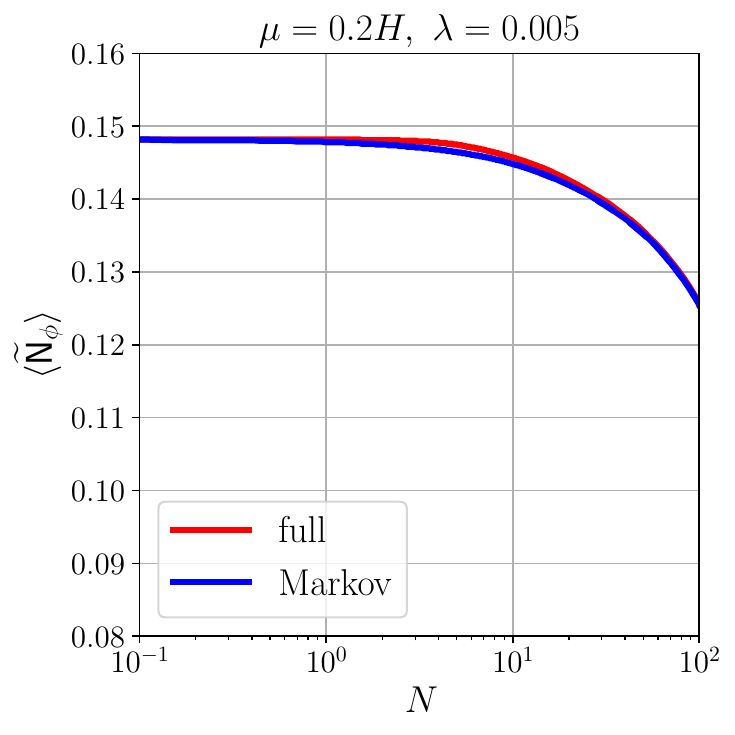}
\end{minipage}\hfill
\begin{minipage}{0.32\columnwidth}
  \centering
  \includegraphics[width=\linewidth]{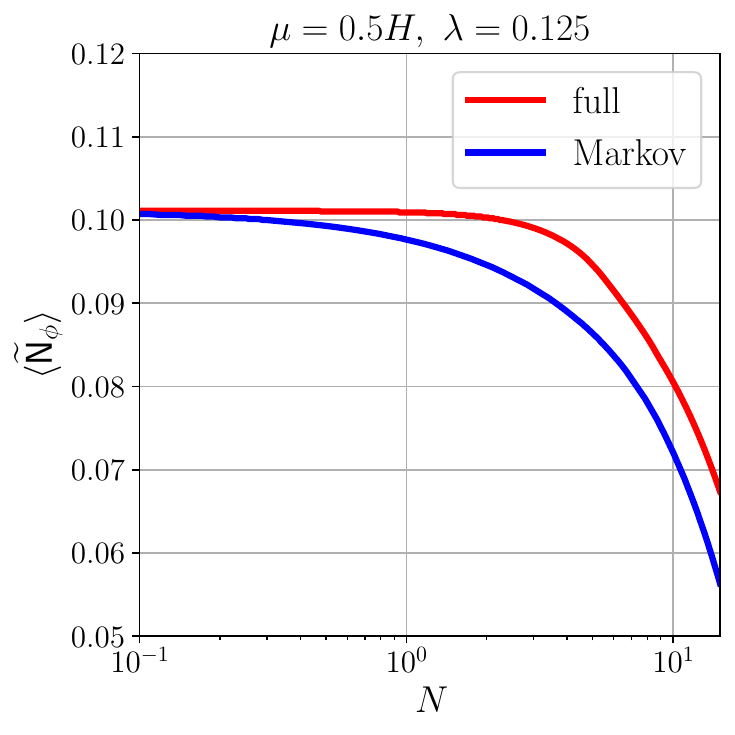}
\end{minipage}

\caption{Second order moments for $\phi$ (upper panels) and the diffusion amplitude for $\phi$ \eqref{normalized noises} in the Langevin equation \eqref{multi phi} (lower panels) for the potential $\mu^2\phi\chi+\lambda\phi^4$ with $\mu=0.2H,\ \lambda=0.125$ (left panels), $\mu=0.2H,\ \lambda=0.005$ (middle panels), and $\mu=0.5H,\ \lambda=0.125$ (right panels). The number of stochastic realizations is $5\times10^4$, and the time step is $\dd N=0.05$.
}
\label{Fig:example21}
\end{figure}

\begin{figure}[h]
\centering

\begin{minipage}{0.32\columnwidth}
  \centering
  \includegraphics[width=\linewidth]{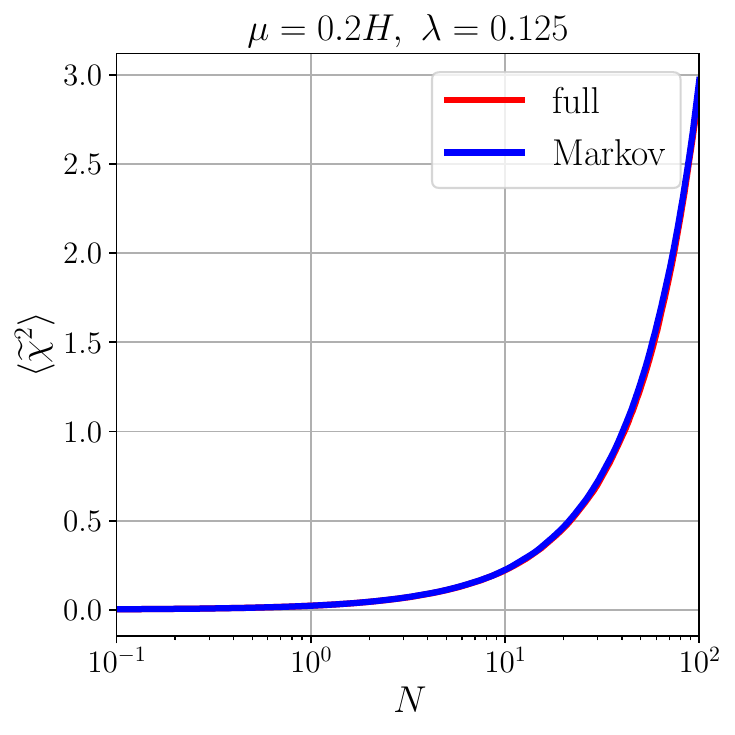}
\end{minipage}\hfill
\begin{minipage}{0.32\columnwidth}
  \centering
  \includegraphics[width=\linewidth]{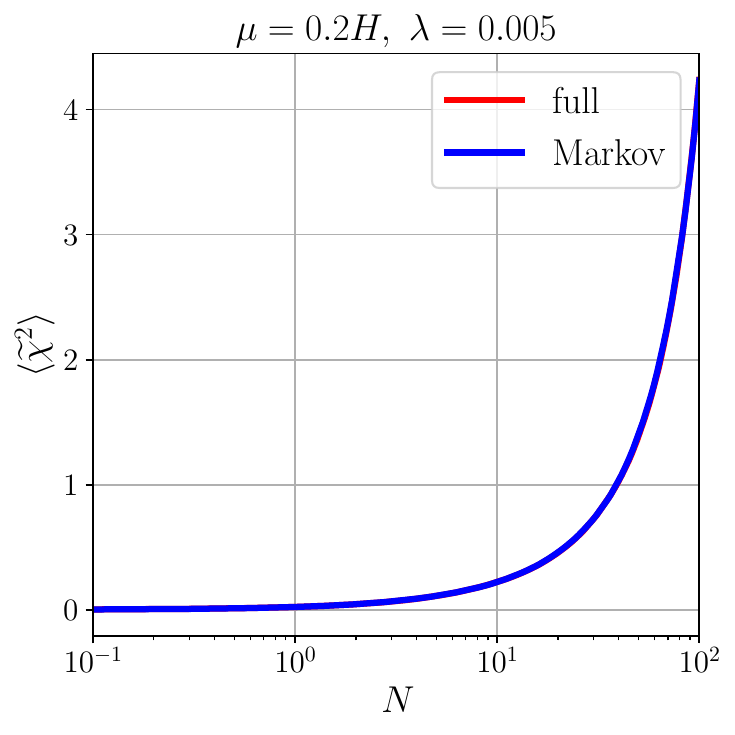}
\end{minipage}\hfill
\begin{minipage}{0.32\columnwidth}
  \centering
  \includegraphics[width=\linewidth]{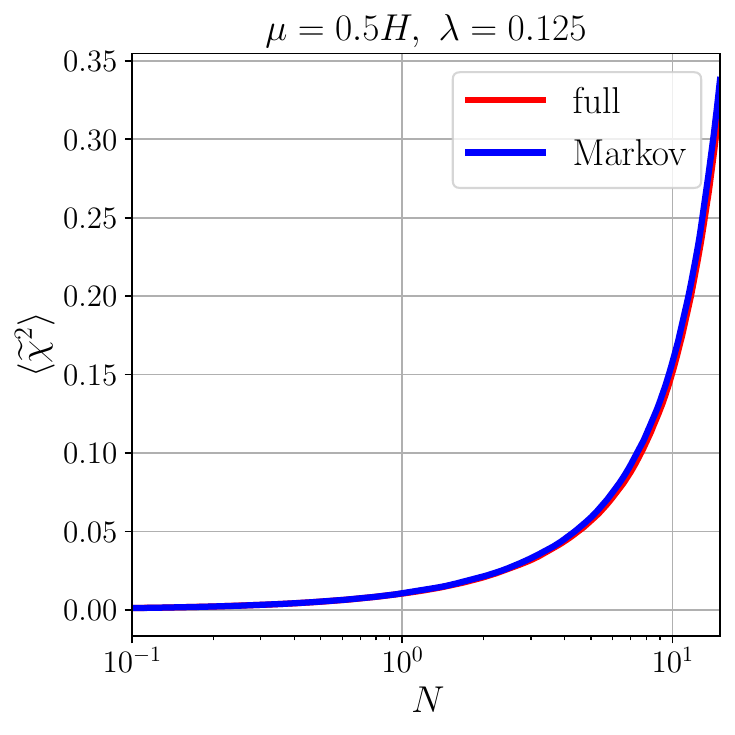}
\end{minipage}

\vspace{2mm}

\begin{minipage}{0.32\columnwidth}
  \centering
  \includegraphics[width=\linewidth]{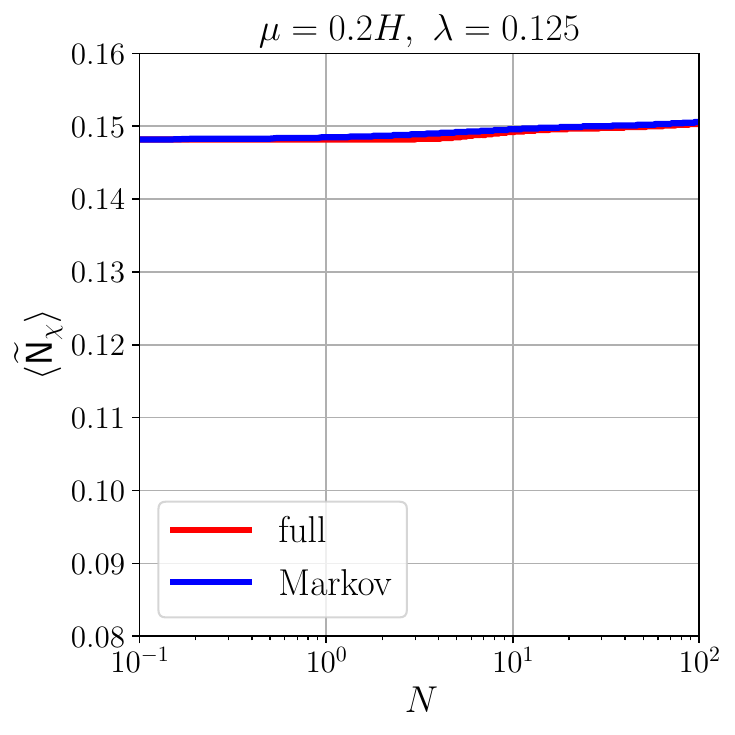}
\end{minipage}\hfill
\begin{minipage}{0.32\columnwidth}
  \centering
  \includegraphics[width=\linewidth]{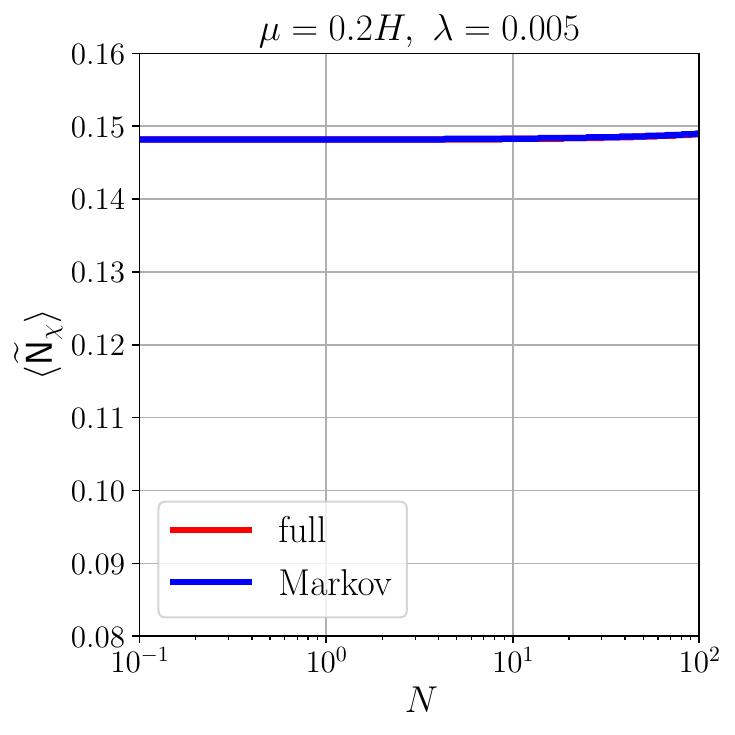}
\end{minipage}\hfill
\begin{minipage}{0.32\columnwidth}
  \centering
  \includegraphics[width=\linewidth]{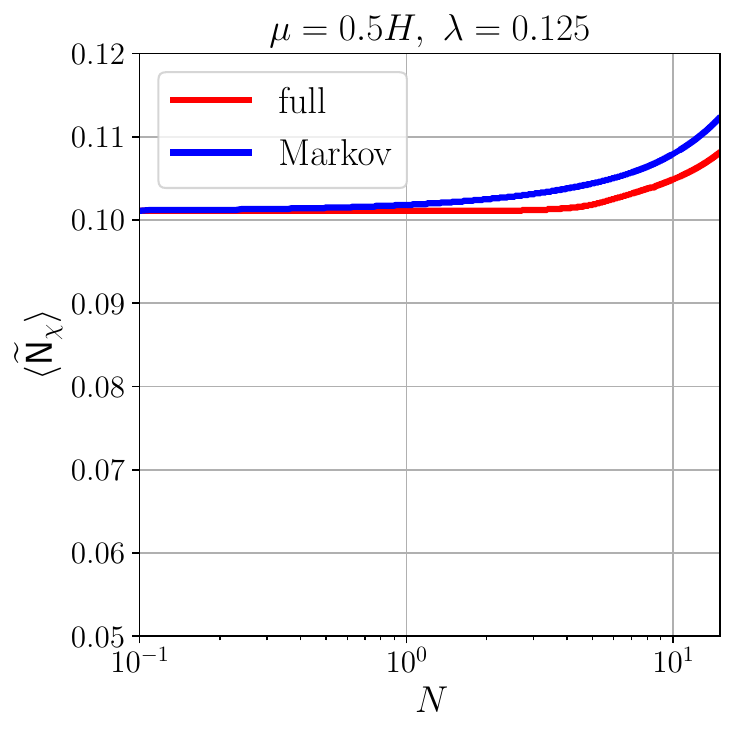}
\end{minipage}

\caption{Second order moments for $\chi$ (upper panels) and the diffusion amplitude for $\chi$ \eqref{normalized noises} in the Langevin equation \eqref{multi chi} (lower panels) for the potential $\mu^2\phi\chi+\lambda\phi^4$ with $\mu=0.2H,\ \lambda=0.125$ (left panels), $\mu=0.2H,\ \lambda=0.005$ (middle panels), and $\mu=0.5H,\ \lambda=0.125$ (right panels). The number of stochastic realizations is $5\times10^4$, and the time step is $\dd N=0.05$.
}
\label{Fig:example22}
\end{figure}

The results of these simulations are presented in \Figs{Fig:example21} and \ref{Fig:example22}. The color plotting follows the same convention as in the previous model. In this multifield case, the massless free fields noise prescription (corresponding to the green curve in the previous subsection) is not meaningful due to the presence of the coupling term through $\mu^2\phi\chi$ in the potential. As a consequence, only the red and blue curves are shown. Broadly speaking, both $\phi$ and $\chi$ initially undergo stochastic diffusion driven by the noises and subsequently grow due to the instability of the potential. This growth persists until additional stabilizing terms in the potential, implicit but omitted in the explicit expression, become dominant.


This potential consists of a bilinear coupling $\mu^2\phi\chi$ and a self-interaction term $\lambda \phi^4$. Hence, the effective mass is independent of $\chi$, and the coupling $\lambda $ controls the degree of non-Markovianity in the same manner as in the previous model, while $\mu$ governs the mixing between the two fields $\phi,\chi$. \Figs{Fig:example21} and~\ref{Fig:example22} illustrate the memory effect: the non-Markovian noise for $\phi$ ($\chi$) decreases (increases) gradually as effective masses develop, reflecting the fact that both fields initially remain close to the origin due to the initial condition \eqref{IC eg2}. In contrast, the Markovian noise, which is evaluated solely from the instantaneous field values, neglects this history of lingering near the origin. As a result, the Markovian treatment overestimates the contributions of the effective masses and yields a smaller diffusion amplitude than the full non-Markovian calculation. Regarding the self-interacting coupling $\lambda$, we find that by comparing the first and second columns in \Figs{Fig:example21} and  \ref{Fig:example22}, the magnitude of non-Markovian effects decreases as the coupling $\lambda$, which efficiently acts as a ``switch" for non-Markovianity, is reduced, in agreement with the behavior observed in the previous model and the general discussion in \Sec{Sec:2}. We also confirm that when this switch is completely turned off ($\lambda=0$), the dynamics of $\phi,\chi$ become equivalent, and their noises are constant since the effective masses are independent of the values of the fields and are then determined solely by $\mu$. 

It can also be confirmed that even though the self-interaction of $\phi$ is the primary source of the discrepancy between the non-Markovian and the Markovian treatment, $\chi$ also acquires the non-Markovian behavior through the coupling $\mu$ to $\phi$, as shown in \Fig{Fig:example22}. Indeed, as $\mu$ increases, the mixing between $\phi$ and $\chi$ becomes stronger; hence, the difference for $\chi$ between the non-Markov and the Markov treatment becomes more pronounced (compare the first and the third columns in \Fig{Fig:example22}). In contrast, as $\mu$ is reduced, the non-Markovianity of $\chi$ diminishes, and the behavior of $\phi$ becomes close to that of the previous model $V=\lambda\phi^4$. 

\section{Summary}\label{Sec:5}

In this work, we have fully solved the systems of the stochastic formalism characterized by both the Langevin equations for IR modes \eqref{phi IR} and \eqref{pi IR} and the equations for UV modes \eqref{phi UV} and \eqref{pi UV} which determine the noise terms in the Langevin equations and are generally required by the derivation of the stochastic formalism. In our numerical simulations, we have applied this framework to three representative models: the MSSM flat direction model \eqref{MSSM potential}, a massless self-interacting single field model with $V=\lambda\phi^4$, and a coupled multifield model with $V=\mu^2\phi\chi + \lambda\phi^4$. We have demonstrated that the full numerical treatment correctly deals with the effective masses and yields a correct phenomenological result for the first model of the MSSM potential in \Sec{Sec:3}. Furthermore, by comparing the full non-Markovian treatment with the Markovian approximation, we have investigated the impact of non-Markovian effects in the single field and the multifield models in \Sec{Sec:4}. 

\Sec{Sec:3} is devoted to the evolution of a flat direction coupled to non-flat directions under stochastic kicks. Our numerical simulation for the MSSM potential \eqref{MSSM potential}, which treats the effective masses unambiguously, demonstrates that the flat direction does not saturate but instead continues to grow freely. This behavior arises because the effective masses of the non-flat directions coupled to the flat direction increase significantly as time evolves. As a result, the non-flat directions, whose field values become small, are unable to suppress the growth of the variance of the flat direction. 
Let us note that the non-Markovianity is also expected to be present in this setup since the system involves strong couplings. However, these effects are not expected to qualitatively alter the overall dynamics; rather, they primarily affect the quantitative details of the behavior, as discussed in \Sec{Sec:4}. Therefore, the essential difference between our work and previous studies \cite{Enqvist:2011pt,Kawasaki:2012bk} lies in the treatment of the contributions from the effective masses.

\Sec{Sec:4} investigates the non-Markovian properties of the systems through a comparison with the Markovian approximation \eqref{Markovphi} and \eqref{Markovpi}. Such non-Markovianity, which is intrinsic to the systems of the stochastic formalism, generally appears because of the scale separation applied to the same fields, as discussed in \Sec{Sec:2}. In the two models considered, $V=\lambda\phi^4$ and $V=\mu^2\phi\chi + \lambda\phi^4$, the memory effect can be directly observed. In particular, setting initial conditions for all fields and momenta to be at the origin, the non-Markovian noises explicitly encode the early-time history during which the system remains localized near the origin. Hence, deviations from the massless free field noises (which neglect effective masses) develop more slowly in the non-Markovian treatment than in the Markovian one. This behavior persists in the multifield case for all fields through the coupling $\mu$ mixing both fields. Another notable feature of non-Markovian noise is that the resulting stationary state differs from that obtained in the Markovian approximation, precisely because of the memory effects. Interestingly, the non-Markovian behavior, consisting of short- and long-term effects, has been found in a different context \cite{Zhang:2012lfb,Ali:2015gac}, even though the equations of interest are not exactly identical. Finally, we have confirmed that the Markovian approximation becomes precisely accurate as the couplings are reduced. This is because, for small couplings, the characteristic time scale of UV modes acting as a bath is much shorter than that of the IR modes as an open system in accordance with the discussion in \Sec{Sec:2}.

Let us comment on the recursive approach to compute the stochastic system~\cite{PerreaultLevasseur:2013eno,PerreaultLevasseur:2013kfq,PerreaultLevasseur:2014ziv,Figueroa:2021zah}. In this approach, the first iteration of the Langevin equations is solved on top of a homogeneous background. Using the statistical quantities obtained in the previous iteration, updated statistical quantities are then computed in the subsequent iteration. Now, let us apply this method to our setup, where all fields and momenta are initialized to zero values. In this case, the homogeneous background solutions coincide with the initial configuration at the origin, which leads to the massless free field noises such as $H/(2\pi)\xi$. However, these noises generically deviate significantly from both the full non-Markovian and Markovian calculations, particularly in multifield cases. Therefore, the first iteration of the recursive approach is generically not enough, even though Ref.~\cite{Figueroa:2021zah} finds it adequate for their specific setup.

Let us note that, in order to reach the above conclusions, several simplifications have been adopted to isolate the physical effects of interest. First, our analysis is performed on a fixed de Sitter background, neglecting metric perturbations. This implies that our results are applicable to test spectator fields or should be regarded as the zero-th-order approximation in a quasi de Sitter expansion. Another point is a bit subtle, mentioned in the footnote \ref{footnote: couplings}. In the formulation of the stochastic formalism, the validity of perturbativity for UV modes is implicitly assumed. From this perspective, the additional correction terms are expected to arise when the coupling becomes stronger. In the present work, however, such corrections are neglected in order to highlight the key physical effects under consideration, namely the treatment of effective masses for the eventual evolution of a flat direction and the role of non-Markovianity. Specifically, the conclusions regarding non-Markovian effects should therefore be interpreted as qualitative benchmarks rather than precise quantitative predictions. A consistent formulation of the stochastic formalism incorporating these correction terms is left for future work.

\section*{Acknowledgments}
T.K. acknowledges financial support from the Institute for Basic Science (IBS) under the project code, IBS-R018-D3 and JST SPRING, Japan Grant Number JPMJSP2180. 
M.K. was supported by JSPS KAKENHI Grant Numbers 25K07297 and by World Premier International Research Center Initiative (WPI), MEXT, Japan.

\appendix

\section{Numerical algorithm for solving systems in the stochastic formalism}\label{Sec:A}
In this appendix, we describe in detail the algorithm employed in the present work. In the stochastic simulation, both the temporal step and the magnitude of the momentum step for UV modes are discretized such that each step in momentum corresponds to a time step. Hence, at each time step, exactly one UV mode crosses the coarse-graining scale $k_\sigma(N)$ on top of de Sitter exponential expansion $a(N)\propto e^N$\footnote{In the simulation, a concrete value of the scale factor $a_\mathrm{ref}$ at a reference time is required. However, the final results do not depend on the absolute value of the scale factor, since only relative values with respect to the reference value are relevant.}. The noises in the Langevin equations are calculated from this mode. The updated IR modes are then obtained and used as input for the UV modes equations at the next time step. At this next time step, the UV mode, just next to the mode that previously exited the coarse-graining scale in the discretized momentum grid, crosses the coarse-graining scale and determines noises for IR modes. This procedure is iterated from the initial time $N=0$. For each stochastic realization, the code executes the algorithm according to the following steps:

\begin{algorithm}
\caption{Numerical solver of the stochastic system for each run $(N\geq0)$}
Set initial conditions $\widetilde\phi^I(0), \widetilde\pi^I(0)$ by hand. \\
Prepare $\phi^I_\uUV(0,k),\pi^I_\uUV(0,k)$ consistent with $\widetilde\phi^I(0), \widetilde\pi^I(0)$ and the BD vacuum \eqref{BD solutions}.\\
\For{$N = 0$ \KwTo $N_{\mathrm{end}}$}{
    Evolve $\widetilde{\phi}^I(N),\widetilde{\pi}^I(N)$ using $\phi^I_\uUV(N,k_\sigma(N)),\pi^I_\uUV(N,k_\sigma(N))$ via Eqs. \eqref{phi IR} and \eqref{pi IR}.\\
    \For{modes $k$ in the simulation $\mathrm{s.t.\ }k_\sigma(N)\leq k \leq k_\uUV(N)$}{
        Evolve $\phi^I_\uUV(N,k),\pi^I_\uUV(N,k)$ using Eqs \eqref{phi UV} and \eqref{pi UV}.}\
}
\KwOut{Statistics of $\widetilde\phi^I(N),\widetilde\pi^I(N),\widetilde{\mathsf{N}}_{\phi^I}(N),\widetilde{\mathsf{N}}_{\pi^I}(N)$.}
\end{algorithm}

Given the initial conditions of the IR modes $\widetilde\phi^I(0), \widetilde\pi^I(0)$, the corresponding UV modes at $N=0$ can be determined, in principle, by imposing the Bunch-Davies vacuum for deep sub-horizon modes:
\begin{align}
    \phi^I_\uUV(N,k)=-i\frac{H}{\sqrt{2k^3}}\frac{k}{aH},\quad\pi^I_\uUV(N,k) = \frac{1}{\sqrt{2k^3}}\prn{\frac{k}{aH}}^{2} \label{BD solutions}
\end{align}
where the phases are chosen as above; other phase choices yield the same observable result, such as a power spectrum. Even though $\widetilde\phi^I(0),\widetilde\pi^I(0)$ can be numerically computed as demonstrated later, in this work we employ the analytical solutions for the UV modes, which is possible due to the simplified initial conditions $\widetilde\phi^I(0)=\widetilde\pi^I(0)=0$. In this case, we solve the UV equations for massless free fields on a fixed de Sitter spacetime, whose analytical solutions are well known.

The Ito-type Langevin equations \eqref{phi IR} and \eqref{pi IR} are solved using the Euler-Maruyama method with a fixed time step $\dd N$. While this approach is straightforward, it has limited precision $\mathcal{O}(\sqrt{\dd N})$ due to the stochastic nature of the noise. Higher order precision methods often require the analytical form of the noises as a function of the IR variables $\widetilde{\mathsf{N}}_{X^I}(\widetilde{\phi}^I(N),\widetilde{\pi}^I(N))$, which cannot be calculated in full non-Markov simulations. 

The equations of UV modes \eqref{phi UV} and \eqref{pi UV} are solved using an explicit Runge-Kutta method of order four with a fixed time step $\dd N$. In practice, the UV cutoff $k_\uUV$ is introduced to avoid numerical divergence from the third term in \eqref{pi UV}. Hence, the UV equations \eqref{phi UV} and \eqref{pi UV} are solved for $k_\sigma(N)\leq k \leq k_\uUV(N)$ where $k_\uUV(N) \equiv c k_\sigma(N)$ with $c\gg1$. The value of $c$ is chosen so that the vacuum oscillation \eqref{BD solutions} remains valid at $k=k_\uUV$ while ensuring sufficient precision of the differential equation solver. Related to this procedure, an $\mathcal{O}(1\%)$ error arises in our results because the time step is not taken to be sufficiently small.

For completeness, we describe the procedure for preparing the UV mode solutions at the initial stage. The stochastic evolution begins at $N=0$, and prior to this time, no stochastic effects are present. Before $N=0$, UV modes evolve from the vacuum oscillation of the Bunch-Davies vacuum \eqref{BD solutions} on top of a deterministic IR background (without stochastic contributions). Given the initial conditions of IR modes $\widetilde\phi^I(0), \widetilde\pi^I(0)$, one can determine the inverse evolution of the deterministic IR modes and then solve the UV mode equations on top of this background. The algorithm to obtain these is as follows:
\begin{algorithm}
\caption{Preparation of UV modes solutions $(N\leq0)$}
Set the IR modes $\widetilde\phi^I(N_*),\widetilde\pi^I(N_*)$ compatible with $\widetilde\phi^I(0), \widetilde\pi^I(0)$ at time $N_*$.\\
\For{modes $k$ in the simulation $\mathrm{s.t.\ }k_\sigma(0)\leq k \leq k_\uUV(0)$}{
    Initialize UV modes $\phi^I_\uUV(N_*,k),\pi^I_\uUV(N_*,k)$ using BD vacuum \eqref{BD solutions}.\\
    \For{$N=N_*$ \KwTo $0$}{
    Evolve $\widetilde\phi^I(N),\widetilde\pi^I(N)$ using Eqs. \eqref{phi IR} and \eqref{pi IR} without noises, and $\phi^I_\uUV(N,k),\pi^I_\uUV(N,k)$ using Eqs. \eqref{phi UV} and \eqref{pi UV}.\ 
    }
}
\KwOut{$\phi^I_\uUV(0,k),\pi^I_\uUV(0,k).$}
\end{algorithm}
Practically, $N_*$ can be chosen appropriately to correspond to the UV cutoff $k_\uUV$.

\section{Complimentary figures}\label{Sec:B}
\begin{figure}[H]
 \begin{center} 
  \subfigure{
   \includegraphics[width=.5\columnwidth]{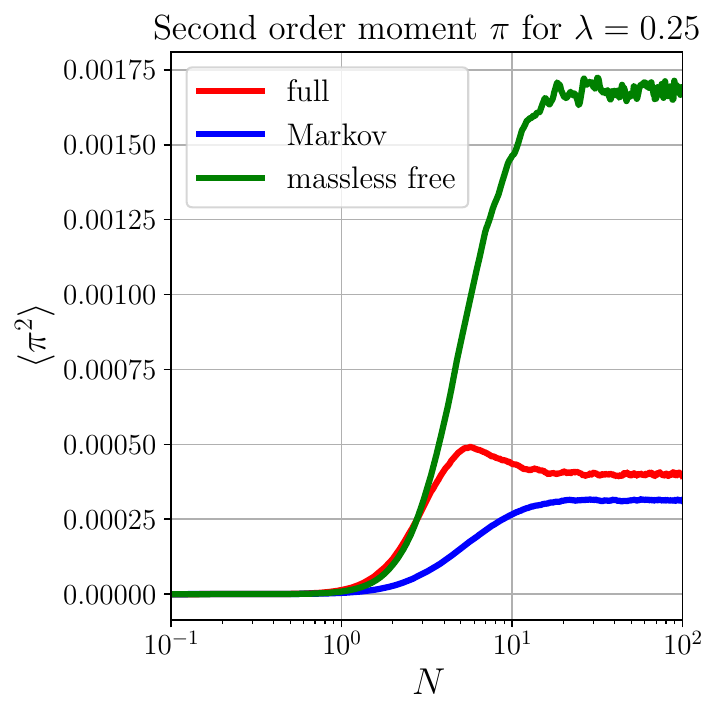}
  }~
  \subfigure{
   \includegraphics[width=.5\columnwidth]{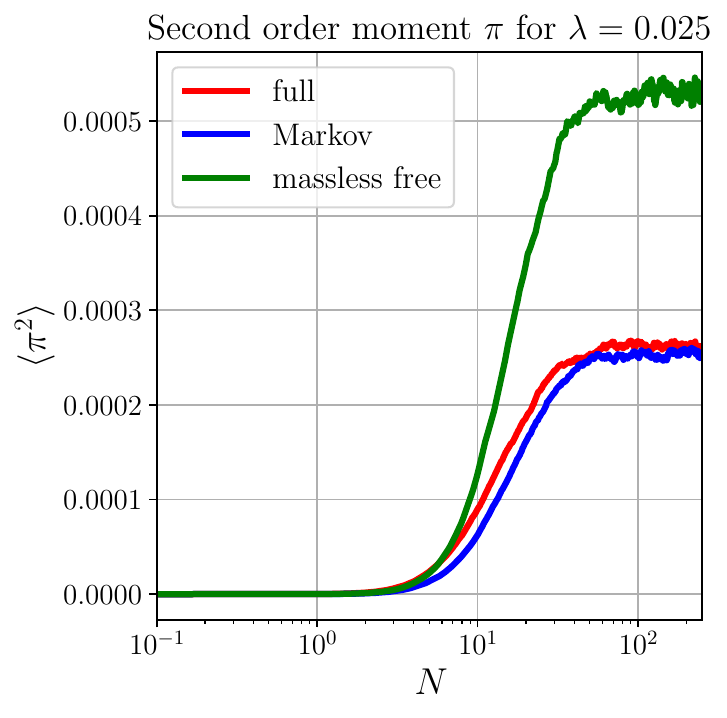}
  }\\
  \subfigure{
   \includegraphics[width=.5\columnwidth]{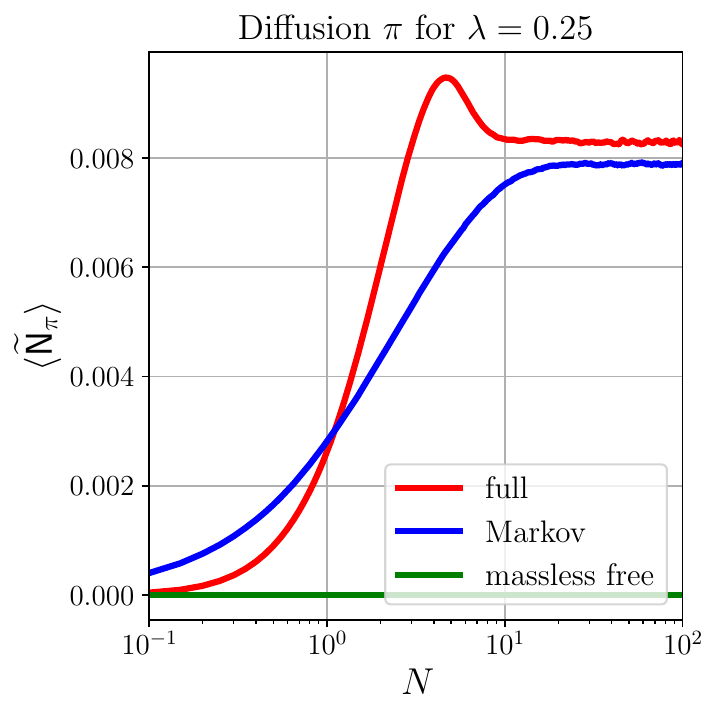}
  }~
  \subfigure{
   \includegraphics[width=.5\columnwidth]{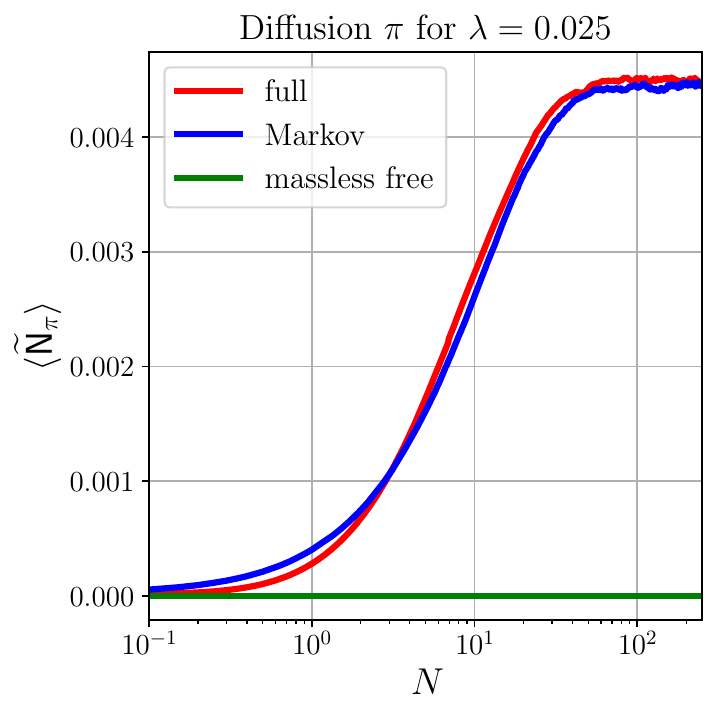}
  }
  \caption{Second order moments (upper panels) and the amplitude of diffusion terms \eqref{normalized noises} in Langevin equations \eqref{single phi} and \eqref{single pi} (lower panels) of the conjugate momentum $\pi$ for the potential $\lambda\phi^4$ with $\lambda=0.25$ (left panels) and $\lambda = 0.025$ (right panels). The number of stochastic realization are $ 1.0\times10^5$ and $5\times 10^4$ for the former and the latter cases. The time step is $\dd N=0.05$.} 
  \label{Fig:example12}
 \end{center}
\end{figure}

This section presents additional figures for the single-field case with $V=\lambda\phi^4$ in \Sec{Sec:42}. In particular, \Fig{Fig:example12} shows the results for the momentum variables, serving as a complement to \Fig{Fig:example11}.

\bibliographystyle{JHEP}
\bibliography{ref}
\end{document}